# Topological Classifications and Bifurcations of Periodic Orbits in the Potential Field of Highly Irregular-shaped Celestial Bodies

Yu Jiang[1,2], Yang Yu[2], Hexi Baoyin[2]

1. State Key Laboratory of Astronautic Dynamics, Xi'an Satellite Control Center, Xi'an 710043, China

2. School of Aerospace Engineering, Tsinghua University, Beijing 100084, China

Y. Jiang (✉) e-mail: jiangyu_xian_china@163.com (corresponding author)

**Abstract.** This paper studies the distribution of characteristic multipliers, the structure of submanifolds, the phase diagram, bifurcations and chaotic motions in the potential field of rotating highly irregular-shaped celestial bodies (hereafter called irregular bodies). The topological structure of the submanifolds for the orbits in the potential field of an irregular body is shown to be classified into 34 different cases, including 6 ordinary cases, 3 collisional cases, 3 degenerate real saddle cases, 7 periodic cases, 7 period-doubling cases, 1 periodic and collisional case, 1 periodic and degenerate real saddle case, 1 period-doubling and collisional case, 1 period-doubling and degenerate real saddle case, and 4 periodic and period-doubling cases. The different distribution of the characteristic multipliers has been shown to fix the structure of the submanifolds, the type of orbits, the dynamical behaviour and the phase diagram of the motion. Classifications and properties for each case are presented. Moreover, tangent bifurcations, period-doubling bifurcations, Neimark-Sacker bifurcations and the real saddle bifurcations of periodic orbits in the potential field of an irregular body are discovered. Submanifolds appear to be Mobius strips and Klein bottles when the period-doubling bifurcation occurs.

# 1. Introduction

The discoveries of large size ratio binary [1, 2] or triple asteroids [3-5] and space missions [6, 7] to minor bodies in the solar system have made the study of nonlinearly dynamical behaviour in the potential field of rotating highly irregular-shaped celestial bodies important [8-10]. This paper aims to discuss the nonlinearly dynamical behaviour of a massless particle in the potential field of irregular bodies, which can be asteroids, comets, or satellites of planets. The theory can be applied to the study of the dynamical behaviour of minor bodies, including a spacecraft around an irregular body and the large size ratio binary or triple asteroids, such as the binary asteroids 1862 Apollo [11], (162000) 1990OS [12], and 41 Daphne [13] and the triple asteroids 87 Sylvia [3] and 216 Kleopatra [5, 14].

Periodic orbits in the potential field of rotating highly irregular-shaped celestial bodies exist and can be found by analytic continuation with respect to certain parameters [9][15-18]. Periodic orbits can be classified into different families using the position characteristics [15], the geometrical appearances [19], and the topological characteristics [18]. Related to the concrete case of finding periodic orbits around rotating highly irregular-shaped celestial bodies, Scheeres et al. [15] calculated periodic orbits about asteroid 4769 Castalia using Poincaré maps and Newton-Raphson iteration and found three main families of periodic orbits close to asteroid Castalia: quasi-equatorial direct, quasi-equatorial retrograde, and non-equatorial. Yu and Baoyin [19] developed a hierarchical searching method to compute periodic orbits around irregular bodies and classified periodic orbits near asteroid 216 Kleopatra into 29 families using the geometrical appearances and an advanced numerical method. Jiang et al. [18] classified periodic orbits near equilibrium points in the potential field of irregular bodies into several families on the basis of topological characteristics.

The study of motion stability in the vicinity of minor irregular bodies includes the stabilities of periodic orbits, quasi-periodic orbits, and equilibrium points. Yu and Baoyin [20] discussed the stabilities of periodic orbits in the potential field of asteroid 216 Kleopatra. The stabilities of equilibrium points determine the motion stabilization near equilibrium points [18][21]. Elipe and Lara [21] modelled the asteroid 433 Eros as a finite straight segment. They found 4 equilibrium points around the segment and discussed their linear stabilities. Mondelo et al. [22] studied asteroid 4 Vesta and found 4 equilibrium points and determined their coordinates and stabilities, showing that two of them are stable, while the other two are unstable. Jiang et al. [18] established the theory about motions near equilibrium points in the potential field of a general rotating irregular body, which includes the linearized motion equation and characteristic equation near equilibrium points, two conditions of stability of the equilibrium points, and discovered that the stabilities of the equilibrium points are determined by the structure of submanifolds and subspaces near the equilibrium points.

Furthermore, bifurcations of motion will occur when the parameters are equal to certain special values. Riaguas et al. [23] discussed bifurcations when considering periodic orbits with parameter variations in the gravity field of a massive straight segment. The motions near 1:1 resonance in the gravity field of a massive straight segment with parameter variations can lead to bifurcations [21]; however, the resonances caused by the 1:1 ratio of the rotational period create equilibrium points [18].

In this paper, the nonlinearly dynamical behaviour of a massless particle in the potential field of an irregular body is studied, including the topological classifications and stabilities of periodic orbits and bifurcations of motion. It is shown that the topological structure of submanifolds for the orbits can be classified into 34 different cases. The different distribution of characteristic multipliers fixes the structure of submanifolds, the type of orbits, the dynamical behaviour and the phase diagram of the motion. The classifications and properties for each case are presented.

Additionally, the theory reveals the mechanism of bifurcations for periodic orbits in the potential field of an irregular body. Period-doubling bifurcations, tangent bifurcations, Neimark-Sacker bifurcations and real saddle bifurcations of periodic orbits are discovered. Motions near bifurcation with parametric variation are sensitive to initial conditions. Additionally, it is found that submanifolds appear to be Mobius strips and Klein bottles when the period-doubling bifurcation occurs.

The theory developed here is applied to the asteroids 216 Kleopatra and 6489 Golevka and the comet 1P/Halley to analyse the dynamical behaviour around these irregular celestial bodies. For the asteroid 6489 Golevka, two periodic orbits, which have entirely similar shapes and belong to different cases, are presented as examples. This result implies that periodic orbits that belong to the different cases might have entirely similar shapes; the essential characteristic of periodic orbits is the topological type rather than shape. The number of different shapes is infinite, while the number of different topological types is finite. For the comet 1P/Halley, two periodic orbits are presented. These two periodic orbits belong to different periodic cases: one belongs to Case P2, the stable periodic case, while the other belongs to Case P4, the unstable periodic case.

Dynamical behaviours and bifurcations around the large size ratio triple asteroid 216 Kleopatra are studied. All four types of bifurcations forecasted herein by the theory are discovered in the potential field of the asteroid 216 Kleopatra. Additionally, the attraction and exclusion effect of critical points in the vicinity of critical points around the asteroid 216 Kleopatra is found.

**2. Topological Structures of Periodic Orbits**

In this section, motion equations, characteristic multipliers of the orbit, and the topological structure of submanifolds will be discussed. We will show that different distributions of characteristic multipliers fix the structure of submanifolds, the type of orbits, the dynamical behaviour, and the phase diagram of the motion.

**2.1 Motion Equations**

Let **r** be the body-fixed vector from the celestial body's centre of mass to the particle, **ω** be the rotational angular velocity of the body relative to the inertial space, and $U(\mathbf{r})$ be the gravitational potential of the body, which can be calculated by the polyhedral model [24, 25]) using radar observation data [26, 27]. The efficient potential $V(\mathbf{r})$ can be defined as [15]

$$V(\mathbf{r}) = -\frac{1}{2}(\boldsymbol{\omega} \times \mathbf{r})(\boldsymbol{\omega} \times \mathbf{r}) + U(\mathbf{r}). \tag{1}$$

The zero-velocity manifold for the particle satisfies [15]

$$V(\mathbf{r}) = H, \tag{2}$$

and the forbidden region satisfies $V(\mathbf{r}) > H$, while the allowable region satisfies $V(\mathbf{r}) \leq H$.

Denoting the generalised momentum as $\mathbf{p} = (\dot{\mathbf{r}} + \boldsymbol{\omega} \times \mathbf{r})$, and the generalised coordinate as $\mathbf{q} = \mathbf{r}$, the Lagrangian and Hamilton functions can then be given by

$$L = \frac{\mathbf{p} \cdot \mathbf{p}}{2} - U(\mathbf{q}) = \frac{1}{2}\dot{\mathbf{q}} \cdot \dot{\mathbf{q}} + \dot{\mathbf{q}} \cdot (\boldsymbol{\omega} \times \mathbf{q}) - V(\mathbf{q}), \tag{3}$$

and

$$H = -\frac{\mathbf{p} \cdot \mathbf{p}}{2} + U(\mathbf{q}) + \mathbf{p} \cdot \dot{\mathbf{q}}. \tag{4}$$

Using the efficient potential, the equation of motion can be written as

$$\ddot{\mathbf{r}} + 2\boldsymbol{\omega} \times \dot{\mathbf{r}} + \frac{\partial V(\mathbf{r})}{\partial \mathbf{r}} = 0. \tag{5}$$

Define

$$\mathbf{z} = [\mathbf{p} \quad \mathbf{q}]^T. \tag{6}$$

Then, the dynamical equation can be expressed in the symplectic form

$$\dot{\mathbf{z}} = \mathbf{F}(\mathbf{z}) = \begin{pmatrix} \mathbf{0} & -\mathbf{I} \\ \mathbf{I} & \mathbf{0} \end{pmatrix} \nabla H(\mathbf{z}), \tag{7}$$

where $\mathbf{I}$ and $\mathbf{0}$ are $3 \times 3$ matrices, and $\nabla H(\mathbf{z}) = \left(\dfrac{\partial H}{\partial \mathbf{p}} \quad \dfrac{\partial H}{\partial \mathbf{q}}\right)^T$ is the gradient of $H(\mathbf{z})$. Defining $\mathbf{J} = \begin{pmatrix} \mathbf{0} & -\mathbf{I} \\ \mathbf{I} & \mathbf{0} \end{pmatrix}$, where $\mathbf{J}$ is a symplectic matrix and $\mathbf{J}\nabla H(\mathbf{z})$ is the Hamiltonian vector field on the symplectic manifold, the dynamical equation in the symplectic form can be rewritten as

$$\mathbf{J}\dot{\mathbf{z}} + \nabla H(\mathbf{z}) = 0. \tag{8}$$

**2.2 Characteristic Multipliers**

Let the orbit $\mathbf{z}(t) = \mathbf{f}(t, \mathbf{z}_0)$ and $\mathbf{f}(0, \mathbf{z}_0) = \mathbf{z}_0$ be the solution of Eq. (8). Defining $\mathbf{f}^t : \mathbf{z} \to \mathbf{f}(t, \mathbf{z})$, the set generated by $\mathbf{f}^t$ and $(\mathbf{f}^t)^{-1}$ is a one-parameter diffeomorphic group, and $\mathbf{f}^t$ is a flow of the dynamical system of the particle in the potential field of a rotating celestial body. On the symplectic manifold $(\mathbf{S}, \Omega)$, the orbit $\mathbf{z}(t) = \mathbf{f}(t, \mathbf{z}_0)$ is denoted as $p$. Here $\Omega$ is a differential 2-form.

Denote $S_p(T)$ as the set of periodic orbits with period $T$. For any periodic orbit $p \in S_p(T)$, consider the matrix $\nabla \mathbf{f} := \dfrac{\partial \mathbf{f}(\mathbf{z})}{\partial \mathbf{z}}$, which is a $6 \times 6$ matrix. The state transition matrix is [19]

$$\Phi(t) = \int_0^t \frac{\partial \mathbf{f}}{\partial \mathbf{z}}(p(\tau)) d\tau. \tag{9}$$

The monodromy matrix of the periodic orbit $p \in S_p$ is

$$M = \Phi(T). \tag{10}$$

Eigenvalues of the monodromy matrix are characteristic multipliers of the periodic orbit, which are fixed by the periodic orbit $p$. The state transition matrix is a symplectic matrix, which means that if $\lambda$ is an eigenvalue of the monodromy matrix, then $\lambda^{-1}$, $\bar{\lambda}$, and $\bar{\lambda}^{-1}$ are also eigenvalues of the state transition matrix; namely, all eigenvalues are likely to have the form 1, $-1$, $\mathrm{sgn}(\alpha) e^{\pm \alpha} (\alpha \in \mathrm{R}, |\alpha| \in (0,1))$, $e^{\pm i\beta}\left(\beta\in(0,\pi)\right)$, and $e^{\pm\sigma\pm i\tau}\left(\sigma,\tau\in\mathrm{R};\sigma>0,\tau\in(0,\pi)\right)$, where $\mathrm{sgn}(\alpha)=\begin{cases}1 & (\text{if } \alpha>0)\\ -1 & (\text{if } \alpha<0)\end{cases}$. Additionally, the periodic orbit has a characteristic multiplier equal to 1, and the multiple-number of the characteristic multipliers 1 is even. The Krein collision means that there are two pairs of eigenvalues that are coincident and in the form of $e^{\pm i\beta}\left(\beta\in(0,\pi)\right)$.

On the sympletic manifold, the periodic orbit can be calculated by the hierarchical grid searching method [19]. The hierarchical parameters are determined by the spherical coordinates of the normal vector of the grid plane. The residual of state [19] is examined for preliminary judgment of periodic orbit as the following equation stated

$$\mathbf{x}_{res}=\mathbf{x}_N-\mathbf{x}_0,$$

where $\mathbf{x}_N$ and $\mathbf{x}_0$ are the initial state and end state, respectively. The gradient direction [19] is determined by the initial states of the former orbit,

$$\mathbf{x}_0^{i+1}=\mathbf{x}_0^{i}+\varepsilon\cdot\boldsymbol{\xi}_2,$$

where $\mathbf{r}_0^i$ and $\dot{\mathbf{r}}_0^i$ are the location and velocity of the $i$-th step,

$$\boldsymbol{\xi}_2=\begin{pmatrix}\boldsymbol{\omega}\times(\boldsymbol{\omega}\times\mathbf{r}_0^i)+\nabla U\\ \dot{\mathbf{r}}_0^i\end{pmatrix}.$$

## 2. 3 Distribution of Eigenvalues, Orbits, and Submanifolds

In this section, we will discuss submanifolds of the symplectic manifold and subspace for the orbit in the potential field of an irregular body, which can help us to determine the motion state, classify the orbits, and analyse the bifurcations. Let

$C_\lambda = \{\lambda \in \mathbb{C} | \lambda \text{ is the characteristic multiplier of the orbit } p\}$, the Jacobian constant be $H = c$, and the eigenvector of the characteristic multiplier $\lambda_j$ be $\mathbf{u}_j$. Denote the tangent space at the point $p$ as $T_p\mathbf{S} = T_p(\mathbf{S}, \Omega)$. Defining $E(p) = span\{\mathbf{u}_j | \lambda_j \in C_\lambda\}$, $E(p)$ is the tangent space of the symplectic manifold $(\mathbf{S}, \Omega)$ at $p$, $T_p\mathbf{S} = E(p)$. $\Omega|_p$ is the non-degenerate anti-symmetric bilinear 2-form on the tangent space $T_p\mathbf{S}$, and $(T_p\mathbf{S}, \Omega|_p) = (E(p), \Omega|_p)$ is a symplectic space.

Denote the asymptotically stable subspace $E^s(p)$, the asymptotically unstable subspace $E^u(p)$, the central subspace $E^c(p)$, the periodic subspace $E^e(p)$, and the periodic doubling subspace $E^d(p)$ as

$$E^s(p) = span\{\mathbf{u}_j | \lambda_j \in C_\lambda, |\lambda_j| < 1\},$$

$$E^c(p) = span\{\mathbf{u}_j | \lambda_j \in C_\lambda, |\lambda_j| = 1\},$$

$$E^u(p) = span\{\mathbf{u}_j | \lambda_j \in C_\lambda, |\lambda_j| > 1\},$$

$$E^e(p) = span\{\mathbf{u}_j | \lambda_j \in C_\lambda, \lambda_j = 1\}, \text{ and}$$

$$E^d(p) = span\{\mathbf{u}_j | \lambda_j \in C_\lambda, \lambda_j = -1\}.$$

Denote the asymptotically stable subspaces $\bar{E}^s(p)$ and $\tilde{E}^s(p)$ as well as the asymptotically unstable subspaces $\bar{E}^u(p)$ and $\tilde{E}^u(p)$ as

$$\begin{cases} \bar{E}^s(p) = span\{\mathbf{u}_j | \lambda_j \in C_\lambda, |\lambda_j| < 1, \operatorname{Im}\lambda_j = 0\} \\ \tilde{E}^s(p) = span\{\mathbf{u}_j | \lambda_j \in C_\lambda, |\lambda_j| < 1, \operatorname{Im}\lambda_j \neq 0\} \end{cases} \text{ and}$$

$$\begin{cases} \bar{E}^u(p) = span\{\mathbf{u}_j | \lambda_j \in C_\lambda, |\lambda_j| > 1, \operatorname{Im}\lambda_j = 0\} \\ \tilde{E}^u(p) = span\{\mathbf{u}_j | \lambda_j \in C_\lambda, |\lambda_j| > 1, \operatorname{Im}\lambda_j \neq 0\} \end{cases}.$$

The asymptotically stable manifold $W^s(\mathbf{S})$, the asymptotically unstable manifold $W^u(\mathbf{S})$, the central manifold $W^c(\mathbf{S})$, the periodic manifold $W^e(\mathbf{S})$, and the periodic doubling manifold $W^d(\mathbf{S})$ of the orbit $p$ (which is seen as a point in the 6-dimensional symplectic manifold) are tangent to the asymptotically stable subspace $E^s(p)$, the asymptotically unstable subspace $E^u(p)$, the central subspace $E^c(p)$, the periodic subspace $E^e(p)$, and the periodic doubling subspace $E^d(p)$, respectively. The asymptotically stable manifolds $\bar{W}^s(\mathbf{S})$ and $\tilde{W}^s(\mathbf{S})$ are tangent to the asymptotically stable subspaces $\bar{E}^s(p)$ and $\tilde{E}^s(p)$, respectively. The asymptotically unstable manifolds $\bar{W}^u(\mathbf{S})$ and $\tilde{W}^u(\mathbf{S})$ are tangent to the asymptotically unstable subspaces $\bar{E}^u(p)$ and $\tilde{E}^u(p)$, respectively.

Denote $\simeq$ as the topological homeomorphism, $\cong$ as the diffeomorphism, and $\oplus$ as the direct sum. Then, $(\mathbf{S}, \Omega) \cong W^s(\mathbf{S}) \oplus W^c(\mathbf{S}) \oplus W^u(\mathbf{S})$ and $T_p \mathbf{S} \cong E^s(p) \oplus E^c(p) \oplus E^u(p)$.

The Krein collision of characteristic multipliers leads to the appearance of the collisional manifold and the collisional subspace. Denote $W^r(\mathbf{S})$ as the collisional manifold, which is tangent to the collisional subspace $E^r(p) = span\{\mathbf{u}_j \mid |\lambda_j| = 1, \operatorname{Im}\lambda_j \neq 0, \exists \lambda_k, s.t. \lambda_k = \lambda_j, j \neq k\}$. Thus, $E^r(p) \subseteq E^c(p)$ and $W^r(\mathbf{S}) \subseteq W^c(\mathbf{S})$. Denote $W^f(\mathbf{S})$ as the uniform manifold, which is tangent to the uniform subspace $E^f(p) = span\{\mathbf{u}_j \mid \lambda_j \in C_\lambda, \exists \lambda_k, s.t. \operatorname{Re}\lambda_j = \operatorname{Re}\lambda_k \neq 0, \operatorname{Im}\lambda_j = \operatorname{Im}\lambda_k = 0, j \neq k\}$. When $\dim E^f(p) \neq 0$, the manifolds and subspaces of $\lambda_j$ and $\lambda_k$ are uniform and the phase diagrams are coincident.

Denote $E^l(p) = span\{\mathbf{u}_j \mid |\lambda_j| = 1, \operatorname{Im}\lambda_j \neq 0, s.t. \forall \lambda_k \in C_\lambda, \lambda_k \neq \lambda_j\}$ as the strongly

stable space, where $J$ is the number of $\lambda_j$ that satisfies $\begin{cases} |\lambda_j| = 1 \\ \text{Im}\,\lambda_j \neq 0 \end{cases}$, and $Z$ is the set of integers. Denote $W^l(\mathbf{S})$ as the strongly stable manifold, which is tangent to the strongly stable subspace and yields $E^c(p) = E^e(p) \oplus E^d(p) \oplus E^r(p) \oplus E^l(p)$ and $W^c(\mathbf{S}) = W^e(\mathbf{S}) \oplus W^d(\mathbf{S}) \oplus W^r(\mathbf{S}) \oplus W^l(\mathbf{S})$. The topological structure of submanifolds fixes the phase diagram of the motion.

Based on the discussion above, one can obtain the following theorem about the topological classification of submanifolds for the periodic orbits in the potential field of a rotating body.

**Theorem 1.** Consider the topological structure of the manifolds for an orbit in the potential field of a rotating celestial body. There exist 6 ordinary cases, 3 purely collisional cases, 3 purely degenerate real saddle cases, 7 purely periodic cases, 7 purely period-doubling cases, 1 periodic and collisional case, 1 periodic and degenerate real saddle case, 1 period-doubling and collision case, 1 period-doubling and degenerate real saddle case, and 4 periodic and period-doubling cases. Distributions of eigenvalues on the complex plane for these cases are shown in Figure 1, and classifications and properties for these cases are shown in Table 1 in Appendix 1. □

The mixed case is a subtype of at least 2 different cases; for example, Case PK1 is a subtype of the periodic cases, meanwhile a subtype of the collisional cases. Consider the orbit as a point in the symplectic manifold. Then, the phase diagram on the asymptotically stable manifold $\bar{W}^s(\mathbf{S})$ is the motion approach to the point as a sink, while the phase diagram on the asymptotically stable manifold $\tilde{W}^s(\mathbf{S})$ is the motion approach to the point as a spiral sink. The phase diagram on the asymptotically unstable manifold $\bar{W}^u(\mathbf{S})$ is the motion leaving the point as a source, while the phase diagram on the asymptotically unstable manifold $\tilde{W}^u(\mathbf{S})$ is the motion leaving the point as a spiral source. The phase diagram on the central manifold $W^c(\mathbf{S})$ is the motion around the point as a centre. Moreover, the phase diagram on the manifold $\bar{W}^s(\mathbf{S}) \oplus \bar{W}^u(\mathbf{S})$ is the motion around the point as a real saddle, while the phase diagram on the manifold $\tilde{W}^s(\mathbf{S}) \oplus \tilde{W}^u(\mathbf{S})$ is the motion around the point as a complex and spiral saddle.

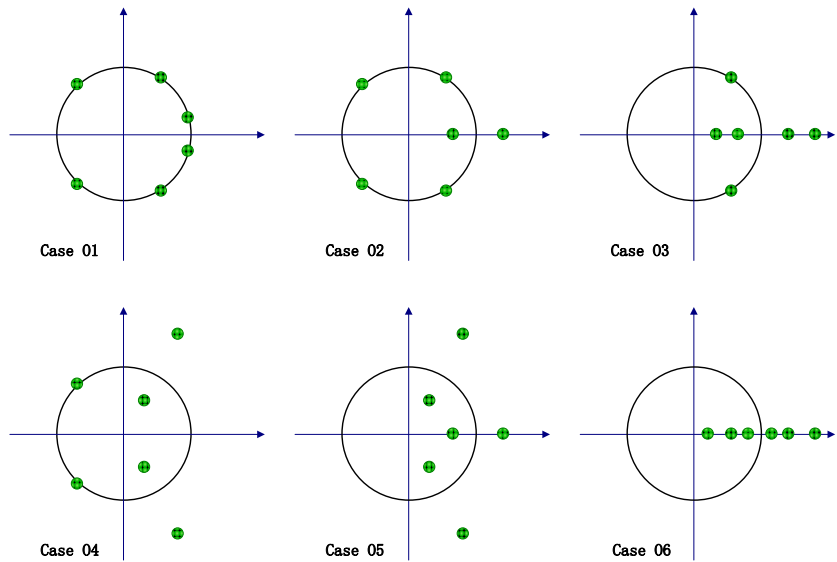

Fig. 1a. Ordinary cases

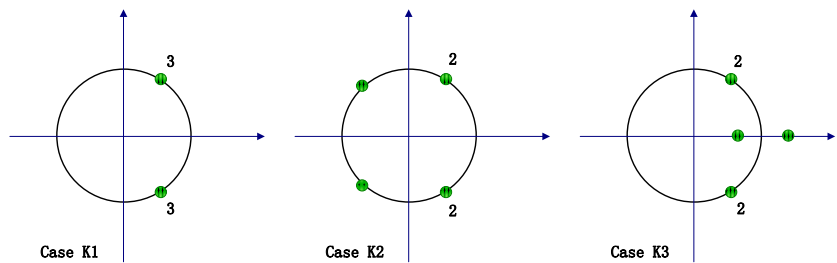

Fig. 1b. Purely collisional cases

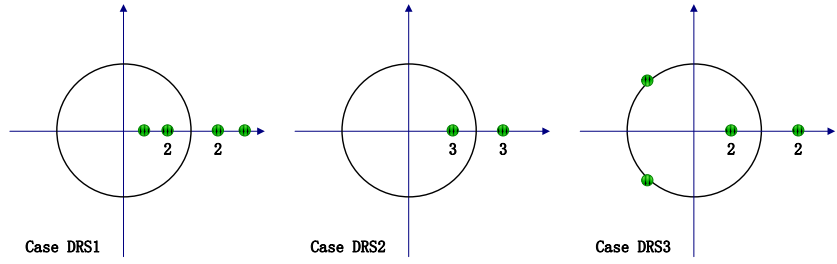

Fig. 1c. Purely degenerate real saddle cases

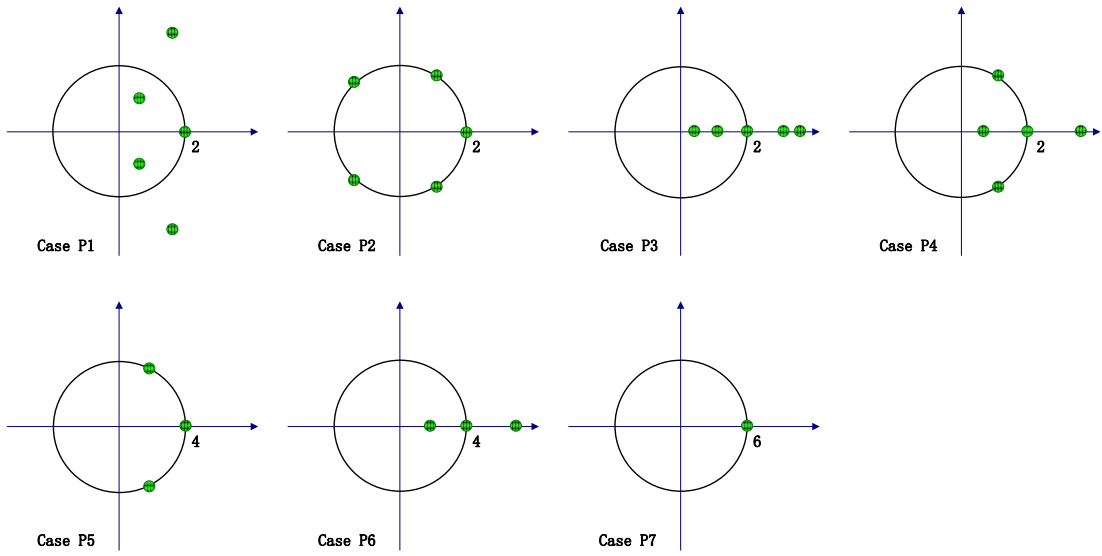

Fig. 1d. Purely periodic cases

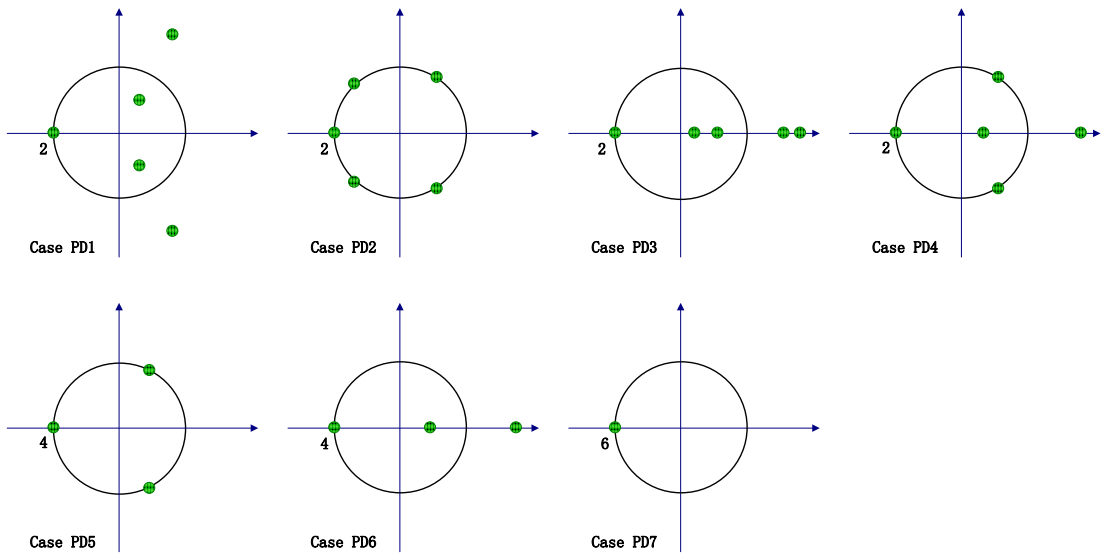

Fig. 1e. Purely period-doubling cases

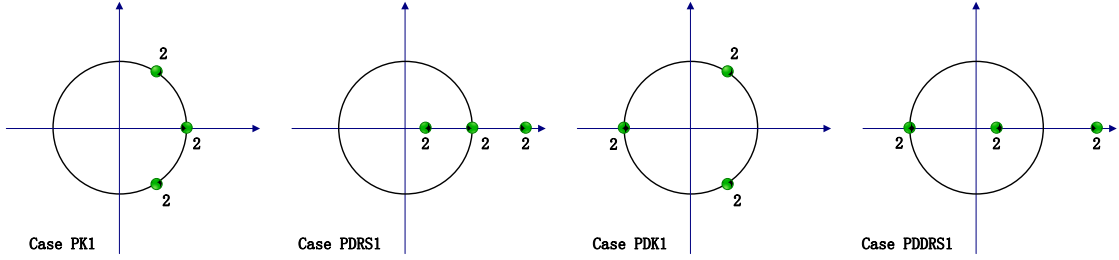

Fig. 1f. Periodic and collisional cases    Fig. 1g. Periodic and degenerate real saddle cases    Fig. 1h. Period-doubling and collisional cases    Fig. 1i. Period-doubling and degenerate real saddle cases

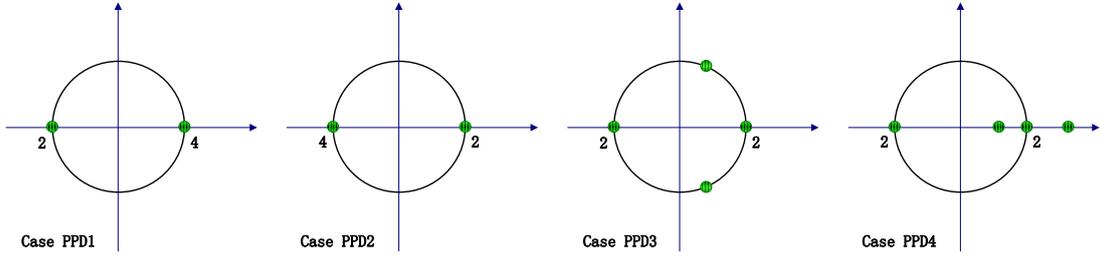

Fig. 1j. Period and period-doubling cases

## 3 Bifurcations of the Periodic Orbits

In this section, we discuss bifurcations of quasi-periodic and periodic orbits in the potential field of an irregular body. The following lemma concerns the motion of characteristic multipliers.

**Lemma 1.** Consider the characteristic multipliers of orbits for the motion in the potential field of an irregular body, that the characteristic multipliers on the unit circle will not leave the unit circle before colliding, and that the collisional point is at $1$, $-1$ or $e^{\pm i\beta}\left(\beta \in (0,\pi)\right)$; the characteristic multipliers on the real axis will not leave the real axis before colliding, and the collisional point is at $1$, $-1$ or $\text{sgn}(\alpha)e^{\pm\alpha}\left(\alpha \in \mathrm{R}, |\alpha| \in (0,1)\right)$.

*Proof.* Let $M$ be the state transition matrix. Then, if $\lambda_0$ is a characteristic multiplier and it is the eigenvalue of $M$, the values $\frac{1}{\lambda_0}$, $\bar{\lambda}_0$, $\frac{1}{\bar{\lambda}_0}$ are also eigenvalues of $M$. Denote $q(\lambda) = \det(M - \lambda I)$, and $q(\lambda)$ is a sextic polynomial. Assume that there exists a single characteristic multiplier $\lambda_0 = e^{i\beta} (\beta \in (0, \pi))$ that will leave the unit circle before colliding when parameters change. Then, $\lambda_1 = \bar{\lambda}_0 = e^{-i\beta} (\beta \in (0, \pi))$ will also leave the unit circle before colliding when parameters change. After leaving the unit circle, the characteristic multipliers satisfy $\lambda_0 = e^{\sigma + i\beta} (\sigma, \beta \in \mathrm{R}; \sigma > 0, \beta \in (0, \pi))$ or $\lambda_0 = e^{-\sigma + i\beta} (\sigma, \beta \in \mathrm{R}; \sigma > 0, \beta \in (0, \pi))$. If $\lambda_0 = e^{\sigma + i\beta} (\sigma, \beta \in \mathrm{R}; \sigma > 0, \beta \in (0, \pi))$, then $e^{-\sigma + i\beta}$, $e^{\sigma - i\beta}$, $e^{-\sigma - i\beta} (\sigma, \beta \in \mathrm{R}; \sigma > 0, \beta \in (0, \pi))$ are also eigenvalues of $M$. In other words, 2 new characteristic multipliers appear with no collision, so the sextic polynomial $q(\lambda)$ has at least 8 roots, which contradicts the fundamental theorem of algebra. Thus, the assumption is incorrect, proving that the characteristic multipliers on the unit circle will not leave the unit circle before colliding. Likewise, the characteristic multipliers on the real axis will not leave the real axis before colliding. For characteristic multipliers on the unit circle, the collisional point is also on the unit circle; for characteristic multipliers on the real axis, the collisional point is also on the real axis. □

Bifurcations of periodic orbits can be analysed using this lemma.

**3.1 The Existence and Continuity of Periodic Orbits with the Parameter Changes**

The continuation of periodic orbits depends on the existence and continuity of periodic orbits with the parameter changes. The motion in the potential field of an irregular-shaped celestial body is a dynamical system that is dependent on the parameters

$$\ddot{\mathbf{r}} + 2\boldsymbol{\omega}(\boldsymbol{\mu}_\omega) \times \dot{\mathbf{r}} + \frac{\partial V(\boldsymbol{\mu}, \mathbf{r})}{\partial \mathbf{r}} = 0, \qquad (11)$$

where $V(\boldsymbol{\mu}, \mathbf{r}) = -\frac{1}{2}(\boldsymbol{\omega}(\boldsymbol{\mu}_\omega) \times \mathbf{r})(\boldsymbol{\omega}(\boldsymbol{\mu}_\omega) \times \mathbf{r}) + U(\boldsymbol{\mu}_U, \mathbf{r})$, $\boldsymbol{\mu}_\omega = \boldsymbol{\mu}_\omega(t)$, $\boldsymbol{\mu}_U = \boldsymbol{\mu}_U(t)$ and $\boldsymbol{\mu} = \boldsymbol{\mu}(t) = [\boldsymbol{\mu}_\omega(t), \boldsymbol{\mu}_U(t)]$ are parameters related to time $t$. $\boldsymbol{\mu}_\omega = \boldsymbol{\mu}_\omega(t)$ is the parameter for the rotational angular velocity $\boldsymbol{\omega}$, while $\boldsymbol{\mu}_U = \boldsymbol{\mu}_U(t)$ is the parameter for the gravitational potential $U(\mathbf{r})$. The periodic orbit $p \in S_p(T)$ is dependent on the parameters $\boldsymbol{\mu}$ and the Jacobian integral $H$. There are several causes that make the parameters $\boldsymbol{\mu}_\omega$ and $\boldsymbol{\mu}_U$ change, such as the YORP effect [28-29], the surface grain motion [30-31], the electrostatic and rotational ejection of dust particles[32], the disintegration of small celestial bodies [33], the energetic collision and disruption of rubble-pile asteroids [34], the continuation of periodic orbit families [15], etc.

The YORP effect, which is the effect of solar radiation on the rotation of asteroids, leads to the change of the rotational speed and the rotational axis [28], i.e. $\boldsymbol{\mu}_\omega$ varies; for instance, the asteroid (54509) 2000 PH5's change in spin rate is (2.0 ± 0.2) × $10^{-4}$ deg/day$^2$ [29]. The surface grain motion [30-31], the electrostatic and rotational ejection of dust particles[32], the disintegration of small celestial bodies [33] and the energetic collision and disruption of rubble-pile asteroids [34] make the parameters $\boldsymbol{\mu}_\omega$ and $\boldsymbol{\mu}_U$ synchronously vary. The continuation of periodic orbit families [15][19-20] leads to the change of the parameter $\boldsymbol{\mu}_U$.

The surface grain motion as well as the electrostatic and rotational ejection of dust particles causes small change of the parameters $\boldsymbol{\mu}_\omega$ and $\boldsymbol{\mu}_U$. The effect of solar radiation and planet's gravitational force on the small celestial body's surface grain leads to the motion of the grain on the surface and the change of the mass distribution, and the charge of the particles and the windmill effect caused by the solar radiation pressure are mainly occurring on the surface of the cometary nucleus. The disintegration of asteroids as well as the energetic collision and disruption of rubble-pile asteroids always cause great change of the parameters $\boldsymbol{\mu}_\omega$ and $\boldsymbol{\mu}_U$, for example, the disintegration of the main-belt asteroid P/2013 R3 produced 10 or more distinct components and a comet-like dust tail, the rotational velocity and the shape of the asteroid have a substantial change [34].

The following theorem concerns the existence and continuity of the periodic orbit when the parameters change.

**Theorem 2.** The periodic orbit belonging to Cases P1, P2, P3, P4, PK1, PDRS1, PPD2, PPD3 or PPD4 is existent and continuous when the parameter changes. In other words, if a periodic orbit $p_0 \in S_p(T)$ belongs to one of the abovementioned cases with parameters $(\boldsymbol{\mu}_0, H_0)$, then there exists an open neighbourhood of $(\boldsymbol{\mu}_0, H_0)$, which can be denoted as $G_N(\boldsymbol{\mu}_0, H_0)$, and $\forall (\boldsymbol{\mu}_1, H_1) \in G_N(\boldsymbol{\mu}_0, H_0)$; there exists a periodic orbit $p_1 \in S_p(T + \Delta T)$ with parameters $(\boldsymbol{\mu}_1, H_1) \in G_N(\boldsymbol{\mu}_0, H_0)$.

*Proof.* For the abovementioned cases, the periodic orbit $p_0 \in S_p(T)$ has only two characteristic multipliers equal to 1. Consider the Poincaré surface of section, which is transverse with $H = H_0$ and the periodic orbit $p_0 \in S_p(T)$; denote the Poincaré mapping as $P(\cdot)$; then, the section is a 4-dimensional surface. Eigenvalues of $\dfrac{\partial P(\cdot)}{\partial \mathbf{z}}$ are characteristic multipliers of the orbit, which can be denoted as $\lambda_j (j = 3, 4, 5, 6; \lambda_j \neq 1)$. Then, apply the implicit function theorem to the Poincaré mapping; for any $(\boldsymbol{\mu}_1, H_1) \in G_N(\boldsymbol{\mu}_0, H_0)$, there exists a periodic orbit $p_1 \in S_p(T + \Delta T)$. Thus, the periodic orbit is existent and continuous when the parameter changes. □

Theorem 2 presents the condition of existence and continuity of periodic orbits with the parameter changes.

## 3.2 Bifurcations of the Periodic Orbits

Bifurcations of periodic orbits in the potential field of an irregular body include tangent bifurcations, period-doubling bifurcations and Neimark-Sacker bifurcations. Tangent bifurcations occur when the orbits have characteristic multipliers that cross 1; the multiple number of 1 is 4 or 6. Period-doubling bifurcations occur when the orbits have characteristic multipliers that cross $-1$; the multiple number of $-1$ is 2 or 4. Neimark-Sacker bifurcations occur when the orbits have two equal characteristic multipliers, which are $e^{i\beta}\left(\beta\in(0,\pi)\right)$ or $e^{-i\beta}\left(\beta\in(0,\pi)\right)$.

Denote $P$ as the Poincaré surface of section. Then, $\dim P = 4$. On the manifold $H = c$, denote $Q_0 = (0,0,0,0)$ as the intersection point from the periodic orbit $p_0 \in S_p(T)$ to the Poincaré surface of section $P$. Then, the asymptotically stable manifold $W^s(p_0)$ and the asymptotically unstable manifold $W^u(p_0)$ of the periodic orbit $p_0$ can be expressed as $W^s(p_0) = \{p : \mathbf{f}(t,p) \to p_0, t \to +\infty\}$ and $W^s(p_0) = \{p : \mathbf{f}(t,p) \to p_0, t \to -\infty\}$. The invariant manifold $W^s(Q_0)$ and $W^u(Q_0)$ satisfy $W^s(Q_0) = W^s(p_0) \cap P$ and $W^u(Q_0) = W^u(p_0) \cap P$, which are the asymptotically stable manifold and the asymptotically unstable manifold of the fixed point $Q_0$.

Let $Q_1 = g(Q_0)$ be the first intersection point from the periodic orbit $p \in S_p(T)$ to the Poincaré surface of section $P$ after $t_0$ that satisfies $Q_1 = Q_0$. Then, if the period $T$ orbit is periodic after rotating the celestial body by $k$ circles, we have $g(Q_0) = P^k(Q_0)$.

### 3.2.1 Period-doubling Bifurcation: Mobius Strips and Klein bottles

Let $P$ be a Poincaré surface of section. The following theorem gives a condition for the period-doubling bifurcation of periodic orbits of a massless test particle in the potential field of an irregular body.

**Theorem 3.** Denote $Q_0 = (0,0,0,0)$ as the intersection point from the periodic orbit $p \in S_p(T)$ to the Poincaré surface of section $P$ at time $t_0$. Let $Q_1 = g(Q_0)$ be the first intersection point from the periodic orbit $p \in S_p(T)$ to the Poincaré surface of section $P$ after $t_0$ that satisfies $Q_1 = Q_0$. The function $g(Q)$ satisfies the following conditions:

a) $Q_0 = g(Q_0)$;

b) The periodic orbit $p \in S_p(T)$ has characteristic multipliers equal to 1 and $-1$ and has no other characteristic multipliers;

c) $\left.\dfrac{d^3 g(g(Q))}{dQ^3}\right|_{Q=Q_0} \neq 0$.

Let $g(\mu, Q)$ be a $\mu-$parameter function that satisfies

d) $\begin{cases} g(\mu, Q_0) = g(Q_0) = Q_0 \\ g(0, Q) = g(Q) \end{cases}$;

e) $\left.\dfrac{\partial g(\mu, Q)}{\partial Q}\right|_{Q=Q_0} = (1+\mu)\mathbf{J}_{4\times 4}$, where $\mathbf{J}_{4\times 4} = diag(1,1,-1,-1)$ or $diag(-1,-1,-1,-1)$ is the $4\times 4$ diagonal matrix.

Then, there exists an open neighbourhood of $(0, Q_0)$, which is denoted as $G$, such that for any $(\mu, Q) \in G$: if $\mu\left[\left.\dfrac{d^3 g(g(Q))}{dQ^3}\right|_{Q=Q_0}\right] > 0$, there is no periodic orbit with minimal period $2T$; if $\mu\left[\left.\dfrac{d^3 g(g(Q))}{dQ^3}\right|_{Q=Q_0}\right] < 0$, then there exists a unique periodic orbit with period $2T$ for any $(\mu, Q) \in G$, which can be denoted as $p^* \in S_p(2T)$

with $(\mu^*, Q^*) \in G$. Additionally, the period $2T$ orbit $p^*$ is asymptotically stable if the periodic orbit $p \in S_p(T)$ is asymptotically unstable; the period $2T$ orbit $p^*$ is asymptotically unstable if the periodic orbit $p \in S_p(T)$ is asymptotically stable.

*Proof.* Here, we only prove the cases where the topological structure belongs to Case PPD1 or PPD2. For other cases, consider the restriction in the submanifold $W^d(\mathbf{S}) \oplus W^e(\mathbf{S})$; one can obtain the conclusion easily.

The period $2T$ orbit satisfies

$$g \circ g(\mu, Q) = g(\mu, g(Q)) = Q \tag{12}$$

$Q = Q_0$ is a root of Eq. (12), with period $T$. The derivatives of $g \circ g(\mu, Q)$ have the form

$$\begin{cases} [g \circ g(\mu, Q)]' = g'(\mu, g(\mu, Q)) g'(\mu, Q) \\ [g \circ g(\mu, Q)]'' = g''(\mu, g(\mu, Q))[g'(\mu, Q)]^2 + g'(\mu, g(\mu, Q)) g''(\mu, Q) \end{cases} \tag{13}$$

and the topological structure belonging to Case PPD1 or PPD2 yields:

$$\begin{cases} [g \circ g(0, Q_0)]' = [g \circ g(Q_0)]' = 1 \\ [g \circ g(0, Q_0)]'' = [g \circ g(Q_0)]'' = 0 \end{cases} \tag{14}$$

Then, the Taylor expansion of $g \circ g(\mu, Q)$ is

$$g \circ g(\mu, Q) = (1+\mu)^2 (Q - Q_0) + \frac{a_2(\mu)}{2}(Q - Q_0)^2 + \frac{a_3(\mu)}{6}(Q - Q_0)^3 + \cdots \tag{15}$$

where $a_2(0) = 0$, $a_3(0) = [g \circ g(Q_0)]''' \neq 0$.

Consider the expression

$$\frac{g \circ g(\mu, Q) - (Q - Q_0)}{Q - Q_0} = \mu(\mu + 2) + \frac{a_2(\mu)}{2}(Q - Q_0) + \frac{a_3(\mu)}{6}(Q - Q_0)^2 + \cdots \tag{16}$$

If $\mu\left[\dfrac{d^3g(g(Q))}{dQ^3}\Big|_{Q=Q_0}\right]>0$, $\mu(\mu+2)$ and $a_3(\mu)$ have the same symbol. Eq. (16) has no solution, which means that there is no period $2T$ orbit with the parameters $(\mu,Q)\in G$.

If $\mu\left[\dfrac{d^3g(g(Q))}{dQ^3}\Big|_{Q=Q_0}\right]<0$, $\mu(\mu+2)$ and $a_3(\mu)$ have different symbols. Then, $\mu(\mu+2)+\dfrac{a_2(\mu)}{2}(Q-Q_0)+\dfrac{a_3(\mu)}{6}(Q-Q_0)^2=0$ has two roots, corresponding to a period $2T$ orbit $(Q^*,g(\mu^*,Q^*))$, and is a solution corresponding to a period $2T$ orbit with the parameters $(\mu,Q)\in G$. In addition, $g\circ g(\mu,Q)-Q=0$ has three solutions, which are $Q_0$, $Q^*$, and $g(\mu^*,Q^*)$.

Condition e yields that $T$ is not the period of this periodic orbit. If $\mu>0$, then $p\in S_p(T)$ is unstable, and if $(1+\mu)^2<1$, then the slope of $g\circ g(\mu,Q)-Q$ is less than zero, so the period $2T$ orbit is stable. Similarly, if $\mu<0$, then $p\in S_p(T)$ is stable, and the period $2T$ orbit is unstable. □

**Remark 1.** Assume that the period $2T$ orbit exists; then, if the period $T$ orbit is periodic after rotating the celestial body on the $k$ circle, the period $2T$ orbit is periodic after rotating the celestial body on the $2k$ circle.

**Remark 2.** The transfers about characteristic multipliers pass through $-1$, leading to the period-doubling bifurcations. For example, for Case PPD1, period-doubling bifurcations include Case P5 → Case PPD1 → Case P6 , Case PD2 → Case PPD1 → Case PD4 , Case PD2 → Case PPD1 → Case PD3 , and Case PPD3 → Case PP → Case PP. Figure 2a shows the appearance of the period-doubling bifurcation generated by Case P2 → Case PPD3 → Case P4.

Denote $\lambda_1 = -1$ as a characteristic multiplier for the periodic orbit $p_0$. Denote $W^c(\mathbf{S})|_{\lambda_1}$ as the submanifold of $W^c(\mathbf{S})$ restricted by $\lambda_1$. Then, $\dim W^c(\mathbf{S})|_{\lambda_1} = 1$. Denote $W^{(c)}(p_0)|_{\lambda_1}$ as the manifold expanded by $\dim W^c(\mathbf{S})|_{\lambda_1} = 1$, from which the expansion makes the periodic orbit $p_0$ from a point to a Jordan curve and causes the dimension of $W^{(c)}(p_0)|_{\lambda_1}$ to become 2. Denote $P_r$ as the projective plane, $M_o$ as the Mobius strip, and $\partial$ as the boundary operator. Denote the manifold generated by $\lambda_1^{-1}$ as $W^{(c)}(p_0)|_{\lambda_1^{-1}}$.

**Corollary 1.** The manifolds $W^{(c)}(p_0)|_{\lambda_1}$ and $W^{(c)}(p_0)|_{\lambda_1^{-1}}$ are both Mobius strips. These two Mobius strips satisfy $W^{(c)}(p_0)|_{\lambda_1} \simeq W^{(c)}(p_0)|_{\lambda_1^{-1}} \simeq M_o$ and $\dim W^{(c)}(p_0)|_{\lambda_1} = \dim W^{(c)}(p_0)|_{\lambda_1^{-1}} = 2$. Additionally, the boundary of the manifold $W^{(u)}(p_0)|_{\lambda_1} \oplus W^{(s)}(p_0)|_{\lambda_1^{-1}}$ is a Klein bottle, which satisfies $\partial\left(W^{(u)}(p_0)|_{\lambda_1} \oplus W^{(s)}(p_0)|_{\lambda_1^{-1}}\right) = 2P_r^2$ and $\dim\left(W^{(u)}(p_0)|_{\lambda_1} \oplus W^{(s)}(p_0)|_{\lambda_1^{-1}}\right) = 3$.

Figure 2a shows the appearance of the period-doubling bifurcation: Case P2 $\to$ Case PPD3 $\to$ Case P4 and the Mobius strip for the period-doubling bifurcation. As shown in the figure 2a, $W^{(c)}(p_0)|_{\lambda_1}$ and $W^{(c)}(p_0)|_{\lambda_1^{-1}}$ are both Mobius strips.

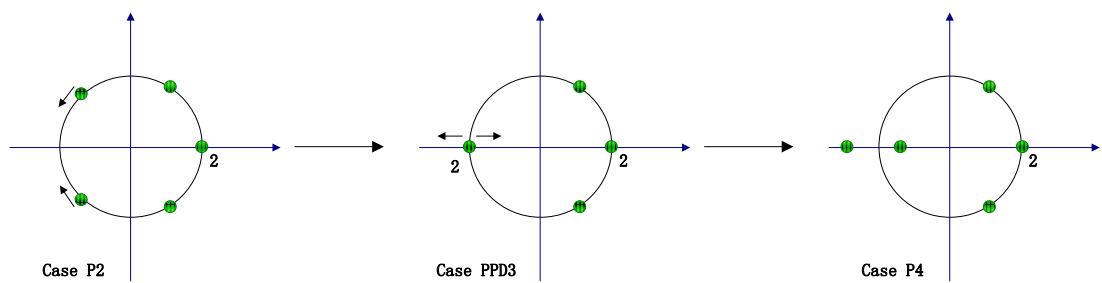

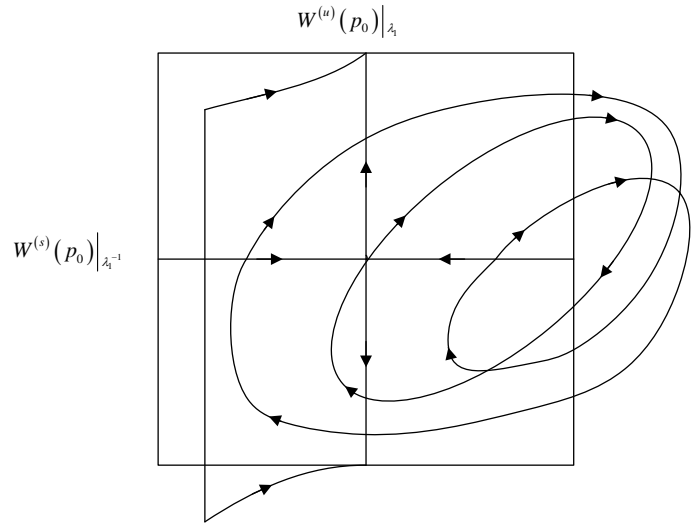

Fig. 2a. Appearance of the period-doubling bifurcation: Case P2 → Case PPD3 → Case P4 and the Mobius strip for the period-doubling bifurcation

### 3.2.2 Tangent Bifurcation

The tangent bifurcation occurs when characteristic multipliers cross the point (1, 0) in the complex plane. The following theorem concerns the appearance of the tangent bifurcation for periodic orbits in the potential field of an irregular body.

**Theorem 4.** Consider the motion of a massless test particle in the potential field of an irregular body. Denote $Q_0 = (0,0,0,0)$ as the intersection point from the periodic orbit $p \in S_p(T)$ to the Poincaré surface of section $P$ at time $t_0$. Let $Q_1 = g(Q_0)$ be the first intersection point from the periodic orbit $p \in S_p(T)$ to the Poincaré surface of section $P$ after $t_0$, satisfying $Q_1 = Q_0$. The function $g(Q)$ satisfies the following conditions:

a) $Q_0 = g(Q_0)$;

b) the periodic orbit $p \in S_p(T)$ has $m$ characteristic multipliers equal to 1, $m = 4$ or $6$.

Let $g(\mu, Q)$ be the $\mu$-parameter function that satisfies

c) $\begin{cases} g(\mu, Q_0) = g(Q_0) = Q_0 \\ g(0, Q) = g(Q) \end{cases}$ ;

d) $\dfrac{\partial g(\mu, Q)}{\partial \mu}\bigg|_{\mu=0, Q=Q_0} \neq 0$ ;

e) $\dfrac{\partial g(\mu, Q)}{\partial Q}\bigg|_{Q=Q_0} = (1+\mu)\mathbf{I}_{4\times 4}$, where $\mathbf{I}_{4\times 4}$ is the $4\times 4$ unit matrix.

Then, there exists an open neighbourhood of $(0, Q_0)$, which is denoted by $G$, such that for any $(\mu, Q) \in G$, if $\dfrac{\partial g(\mu, Q)}{\partial \mu} \cdot \dfrac{\partial^2 g(\mu, Q)}{\partial Q^2} < 0$, then there are no periodic orbits for $\mu < 0$, while there are two periodic orbits for $\mu > 0$; if $\dfrac{\partial g(\mu, Q)}{\partial \mu} \cdot \dfrac{\partial^2 g(\mu, Q)}{\partial Q^2} > 0$, then there are no periodic orbits for $\mu > 0$, while there are two periodic orbits for $\mu < 0$.

*Proof:* The periodic orbit is a fixed point on the Poincaré surface of section $P$. Denote $\delta(\mu, Q) = g(\mu, Q) - Q$, then $\delta(\mu, Q) = 0$ is established if and only if $Q$ is a fixed point for the function $g(\mu, Q)$, and

$$\dfrac{\partial \delta(0, Q_0)}{\partial \mu} = \dfrac{\partial \delta(\mu, Q)}{\partial \mu}\bigg|_{\mu=0, Q=Q_0} = \dfrac{\partial g(\mu, Q)}{\partial \mu}\bigg|_{\mu=0, Q=Q_0} \neq 0$$

Using the implicit function theorem, there exists an open neighbourhood of $(0, Q_0)$, $G$, and there exists a function $p$ that is defined in $G$, such that for any $(\mu, Q) \in G$, $\delta(\mu, Q) = 0$ implies $\mu = p(Q)$, while $\mu = p(Q)$ implies $\delta(\mu, Q) = 0$. Clearly, $0 = p(Q_0)$.

Consider the function $\delta(p(Q), Q) = 0$ at $(\mu, Q) = (0, Q_0)$. Then,

$$\left(\dfrac{\partial \delta(\mu, Q)}{\partial \mu}\bigg|_{\mu=0, Q=Q_0}\right) \cdot \left(p'(Q)\big|_{Q=Q_0}\right) + \left(\dfrac{\partial \delta(\mu, Q)}{\partial Q}\bigg|_{\mu=0, Q=Q_0}\right) = 0$$

Using conditions b and d yields $\left.\dfrac{\partial \delta(\mu,Q)}{\partial \mu}\right|_{\mu=0,Q=Q_0} = \left.\dfrac{\partial g(\mu,Q)}{\partial \mu}\right|_{\mu=0,Q=Q_0} \neq 0$ and $\left.\dfrac{\partial \delta(\mu,Q)}{\partial Q}\right|_{\mu=0,Q=Q_0} = \left.\dfrac{\partial g(\mu,Q)}{\partial Q}\right|_{\mu=0,Q=Q_0} - 1 = 0$. Thus, $\left.p'(Q)\right|_{Q=Q_0} = 0$.

Additionally,

$$\left.\dfrac{\partial^2 g(\mu,Q)}{\partial Q^2}\right|_{\mu=0,Q=Q_0} + \left[\left.\dfrac{\partial^2 g(\mu,Q)}{\partial \mu^2}\right|_{\mu=0,Q=Q_0} + \left.\dfrac{\partial^2 g(\mu,Q)}{\partial \mu \partial Q}\right|_{\mu=0,Q=Q_0}\right] p'(Q)|_{Q=Q_0}$$

$$+ \left[\left.\dfrac{\partial g(\mu,Q)}{\partial \mu}\right|_{\mu=0,Q=Q_0}\right] p''(Q)|_{Q=Q_0} = 0$$

Thus,

$$p''(Q)|_{Q=Q_0} = -\dfrac{\left.\dfrac{\partial^2 g(\mu,Q)}{\partial Q^2}\right|_{\mu=0,Q=Q_0}}{\left.\dfrac{\partial g(\mu,Q)}{\partial \mu}\right|_{\mu=0,Q=Q_0}}$$

Considering $0 = p(Q_0)$, using the Taylor expansion,

$$\mu = p''(Q)\dfrac{Q^2}{2} = -\left(\dfrac{\dfrac{\partial^2 g(\mu,Q)}{\partial Q^2}}{\dfrac{\partial g(\mu,Q)}{\partial \mu}}\right)\dfrac{Q^2}{2}$$

Thus, if $\begin{cases}\mu < 0 \\ \dfrac{\partial g(\mu,Q)}{\partial \mu} \cdot \dfrac{\partial^2 g(\mu,Q)}{\partial Q^2} < 0\end{cases}$ or $\begin{cases}\mu > 0 \\ \dfrac{\partial g(\mu,Q)}{\partial \mu} \cdot \dfrac{\partial^2 g(\mu,Q)}{\partial Q^2} > 0\end{cases}$, the above equation has no solution and there are no periodic orbits; if $\begin{cases}\mu > 0 \\ \dfrac{\partial g(\mu,Q)}{\partial \mu} \cdot \dfrac{\partial^2 g(\mu,Q)}{\partial Q^2} < 0\end{cases}$ or $\begin{cases}\mu > 0 \\ \dfrac{\partial g(\mu,Q)}{\partial \mu} \cdot \dfrac{\partial^2 g(\mu,Q)}{\partial Q^2} < 0\end{cases}$, the above equation has two solutions and there are two periodic orbits. □

The following remark means that the motion state near the tangent bifurcation with parametric variation is sensitive to initial conditions.

**Remark 3.** The transfers about characteristic multipliers pass through 1, leading to tangent bifurcations. For example, for Case P5, tangent bifurcations include Case P2 → Case P5 → Case P4 and Case P4 → Case P5 → Case P2. Figure 2b shows the appearance of the tangent bifurcation generated by Case P2 → Case P4 → Case P3.

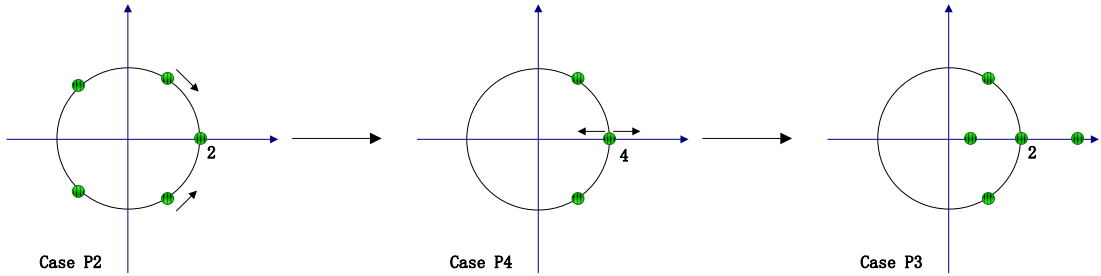

Fig. 2b. Appearance of the tangent bifurcation: Case P2 → Case P4 → Case P3

### 3.2.3 Krein collision and Neimark-Sacker Bifurcation

The following theorem gives a condition for the Neimark-Sacker bifurcation of periodic orbits around an irregular body. The Neimark-Sacker bifurcation occurs when the Krein collision appears.

**Theorem 5.** Consider the motion of a massless test particle in the potential field of an irregular body. Denote $Q_0 = (0,0,0,0)$ as the intersection point from the periodic orbit $p \in S_p(T)$ to the Poincaré surface of section $P$ at time $t_0$. Let $Q_1 = g(Q_0)$ be the first intersection point from the periodic orbit $p \in S_p(T)$ to the Poincaré surface of section $P$ after $t_0$ that satisfies $Q_1 = Q_0$. The function $g(Q)$ satisfies the following conditions:

a) $Q_0 = g(Q_0)$;

b) the periodic orbit $p \in S_p(T)$ has characteristic multipliers equal to 1, $e^{i\beta} (\beta \in (0,\pi))$, and $e^{-i\beta} (\beta \in (0,\pi))$; the multiple number of the characteristic multipliers is equal to 2;

c) $\forall k \in N$, $e^{ik\beta}(\beta \in (0,\pi)) \neq 1$.

Let $g(\mu, Q)$ be the $\mu$-parameter function that satisfies the following:

d) $\begin{cases} g(\mu, Q_0) = g(Q_0) = Q_0 \\ g(0, Q) = g(Q) \end{cases}$;

e) $\left.\dfrac{\partial g(\mu, Q)}{\partial \mu}\right|_{\mu=0, Q=Q_0} \neq 0$;

f) $\left.\dfrac{\partial g(\mu, Q)}{\partial Q}\right|_{Q=Q_0} = (1+\mu)\mathbf{K}_{4\times 4}$, where $\mathbf{K}_{4\times 4} = diag(e^{i\beta}, e^{i\beta}, e^{-i\beta}, e^{-i\beta})$ is the

4×4 diagonal matrix in which $\beta \in (0, \pi)$.

Then,

1) for any $(0, Q_1) \in G$, there exists $(\mu_1, 0) \in G$ and a function $p$ that satisfies $\mu_1 = p(Q_1)$ and corresponds to a periodic orbit; for any $(\mu_1, 0) \in G$, there exists $(0, Q_1) \in G$ and a function $p$ that satisfies $\mu_1 = p(Q_1)$ and corresponds to a periodic orbit;

2) there exists an open neighbourhood of $(0, Q_0)$, which is denoted as $G$, such that for any $(\mu, Q) \in G$, one of the following conditions is established: a) $g(\mu, Q)$ corresponds to an unstable orbit; b) $g(\mu, Q)$ corresponds to an stable quasi-periodic orbit, and the intersection point from the quasi-periodic orbit to the Poincaré surface is on a 2-dimensional tori $T^2$; or c) $g(\mu, Q)$ corresponds to an unstable and collisional quasi-periodic orbit, and the intersection point from the quasi-periodic orbit to the Poincaré surface is on a 1-dimensional tori $T^1$.

*Proof:* Condition e yields that the eigenvalues of the monodromy matrix of the orbit corresponding to $g(\mu, Q)$ are different from the eigenvalues of the monodromy matrix $M$, where $M$ is the monodromy matrix of the periodic orbit $p \in S_p(T)$. The periodic orbit is a fixed point on the Poincaré surface of section $P$. Let $\delta(\mu,Q) = g(\mu,Q) - Q$. Then, $\delta(0,Q_0) = 0$, $\delta(\mu,Q) = 0$ is established if and only if $Q$ is a fixed point for the function $g(\mu,Q)$, and

$$\frac{\partial \delta(0,Q_0)}{\partial \mu} = \frac{\partial \delta(\mu,Q)}{\partial \mu}\bigg|_{\mu=0,Q=Q_0} = \frac{\partial g(\mu,Q)}{\partial \mu}\bigg|_{\mu=0,Q=Q_0} \neq 0$$

Using the implicit function theorem, there exists an open neighbourhood of $(0,Q_0)$, which is denoted by $G$, and there exists a function $p$ that is defined in $G$ such that for any $(\mu,Q) \in G$, $\delta(\mu,Q) = 0 \Leftrightarrow \mu = p(Q)$. In addition, for any $(0,Q_1) \in G$, there exists $(\mu_1,0) \in G$ and a function $p$ that satisfies $\mu_1 = p(Q_1)$ and corresponds to a periodic orbit; for any $(\mu_1,0) \in G$, there exists $(0,Q_1) \in G$ and a function $p$ that satisfies $\mu_1 = p(Q_1)$ and corresponds to a periodic orbit, which leads to conclusion 1.

Consider the topological classification of six characteristic multipliers for the orbit on the complex plane. If at least one characteristic multiplier lies outside the unit circle, $g(\mu,Q)$ corresponds to an unstable orbit. If all of the characteristic multipliers lie on the unit circle and the characteristic multipliers are in the form $(1,1,e^{i\beta},e^{i\beta},e^{-i\beta},e^{-i\beta})(\beta \in (0,\pi))$, the intersection point from the quasi-periodic orbit to the Poincaré surface is on a 1-dimensional tori $T^1$ and $g(\mu,Q)$ corresponds to an unstable and collisional quasi-periodic orbit. Additionally, if all of the characteristic multipliers lie on the unit circle and the characteristic multipliers are in the form $(1,1,e^{i\beta_1},e^{i\beta_2},e^{-i\beta_1},e^{-i\beta_2})(\beta_1,\beta_2 \in (0,\pi))$, the intersection point from the quasi-periodic orbit to the Poincaré surface is on a 2-dimensional tori $T^2$ and

$g(\mu,Q)$ corresponds to a stable quasi-periodic orbit, which leads to conclusion 2. □

**Remark 4.** The transfers about the characteristic multipliers pass through the unit circle without 1 or $-1$, leading to the Neimark-Sacker bifurcations. For example, for Case PK1, the Neimark-Sacker bifurcations include Case P2 → Case PK1 → Case P1 and Case P1 → Case PK1 → Case P2. Figure 2c shows the appearance of the Neimark-Sacker bifurcation generated by Case P2 → Case PK1 → Case P1.

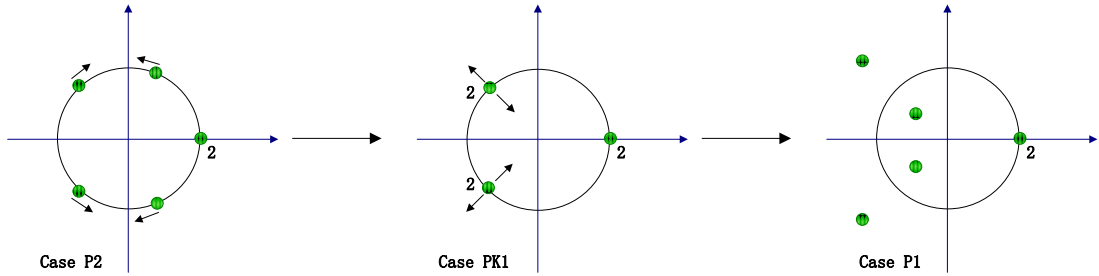

Fig. 2c. Appearance of the Krein collision and Neimark-Sacker bifurcation:
Case P2 → Case PK1 → Case P1

### 3.2.4 Real Saddle Bifurcation

The following theorem gives a condition for the real saddle bifurcation of periodic orbits around an irregular body. Near the real saddle bifurcation, there exists motion of both types of saddle, namely the real saddle and the complex saddle.

**Theorem 5.** Consider the motion of a massless test particle in the potential field of an irregular body. Denote $Q_0 = (0,0,0,0)$ as the intersection point from the periodic orbit $p \in S_p(T)$ to the Poincaré surface of section $P$ at time $t_0$. Let $Q_1 = g(Q_0)$ be the first intersection point from the periodic orbit $p \in S_p(T)$ to the Poincaré surface of section $P$ after $t_0$ satisfying $Q_1 = Q_0$. The function $g(Q)$ satisfies the following conditions:

a) $Q_0 = g(Q_0)$;

b) the periodic orbit $p \in S_p(T)$ has characteristic multipliers equal to 1,

$\text{sgn}(\alpha_j)e^{\alpha_j}$ $(\alpha_j \in \mathbb{R}, |\alpha_j| \in (0,1), j=1)$ , and

$\text{sgn}(\alpha_j)e^{-\alpha_j}$ $(\alpha_j \in \mathbb{R}, |\alpha_j| \in (0,1), j=1)$, where $\text{sgn}(\alpha) = \begin{cases} 1 & (\text{if } \alpha > 0) \\ -1 & (\text{if } \alpha < 0) \end{cases}$ and

the multiple number of the characteristic multipliers is equal to 2.

Let $g(\mu, Q)$ be the $\mu$-parameter function that satisfies

c) $\begin{cases} g(\mu, Q_0) = g(Q_0) = Q_0 \\ g(0, Q) = g(Q) \end{cases}$;

d) $\dfrac{\partial g(\mu, Q)}{\partial \mu}\bigg|_{\mu=0, Q=Q_0} \neq 0$;

e) $\dfrac{\partial g(\mu, Q)}{\partial Q}\bigg|_{Q=Q_0} = (1+\mu)\mathbf{K}_{4\times 4}$, where

$\mathbf{K}_{4\times 4} = diag\left(\text{sgn}(\alpha_j)e^{\alpha_j}, \text{sgn}(\alpha_j)e^{\alpha_j}, \text{sgn}(\alpha_j)e^{-\alpha_j}, \text{sgn}(\alpha_j)e^{-\alpha_j}\right)$ is the $4\times 4$ diagonal matrix in which $|\alpha_j| \in (0,1)$.

Then, we can conclude the following:

1) For any $(0, Q_1) \in G$, there exists $(\mu_1, 0) \in G$ and a function $p$ that satisfies $\mu_1 = p(Q_1)$ and corresponds to an unstable periodic orbit; for any $(\mu_1, 0) \in G$, there exists $(0, Q_1) \in G$ and a function $p$ that satisfies $\mu_1 = p(Q_1)$ and corresponds to an unstable periodic orbit;

2) there exists an open neighbourhood of $(0, Q_0)$, which is denoted by $G$, such that for any $(\mu, Q) \in G$, one of the following conditions is established: a) the topological structure of submanifolds is $(\mathbf{S}, \Omega) \cong W^e(\mathbf{S}) \oplus \tilde{W}^s(\mathbf{S}) \oplus \tilde{W}^u(\mathbf{S})$, and the dimensions of submanifolds satisfy $\dim W^e(\mathbf{S}) = \dim \tilde{W}^s(\mathbf{S}) = \dim \tilde{W}^u(\mathbf{S}) = 2$; b) the topological structure of the submanifold is $(\mathbf{S}, \Omega) \cong W^e(\mathbf{S}) \oplus \bar{W}^s(\mathbf{S}) \oplus \bar{W}^u(\mathbf{S})$, and the dimensions of submanifolds satisfy $\dim W^e(\mathbf{S}) = \dim \bar{W}^s(\mathbf{S}) = \dim \bar{W}^u(\mathbf{S}) = 2$. □

**Remark 5.** The transfers about characteristic multipliers pass through the real axis without 1 or $-1$, leading to the real saddle bifurcations. For example, for Case PDRS1, real saddle bifurcations include Case P1 $\rightarrow$ Case PDRS1 $\rightarrow$ Case P3 and Case P3 $\rightarrow$ Case PDRS1 $\rightarrow$ Case P1. Figure 2d shows the appearance of the real saddle bifurcation generated by Case P1 $\rightarrow$ Case PDRS1 $\rightarrow$ Case P3.

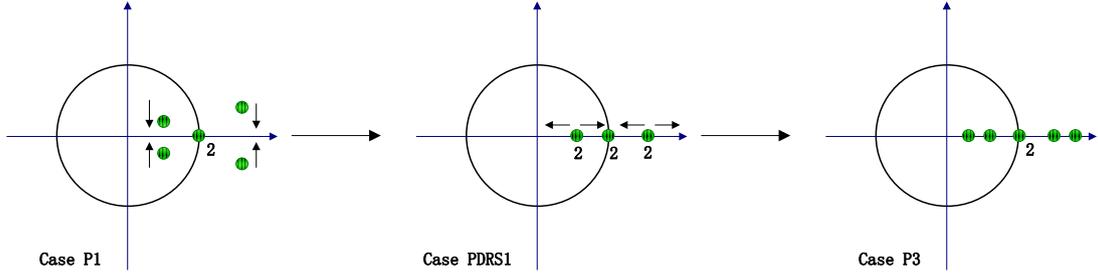

Fig. 2d. Appearance of the real saddle bifurcation: Case P1 $\rightarrow$ Case PDRS1 $\rightarrow$ Case P3

## 4. Applications to Irregular Celestial Bodies

In this section, the theory developed in the previous sections is applied to asteroid 216 Kleopatra and 6489 Golevka, as well as the comet nucleus of 1P/Halley. Physical models of these celestial bodies were calculated using the polyhedral model [24, 25] using data from radar observations [26, 27]. Using the polyhedron method, the gravitational potential of the asteroid [24, 25] can be calculated by

$$U = \frac{1}{2} G\sigma \sum_{e \in edges} \mathbf{r}_e \cdot \mathbf{E}_e \cdot \mathbf{r}_e \cdot L_e - \frac{1}{2} G\sigma \sum_{f \in faces} \mathbf{r}_f \cdot \mathbf{F}_f \cdot \mathbf{r}_f \cdot \omega_f . \quad (17)$$

Besides[25],

$$\nabla U = -G\sigma \sum_{e \in edges} \mathbf{E}_e \cdot \mathbf{r}_e \cdot L_e + G\sigma \sum_{f \in faces} \mathbf{F}_f \cdot \mathbf{r}_f \cdot \omega_f , \quad (18)$$

$$\nabla(\nabla U) = G\sigma \sum_{e \in edges} \mathbf{E}_e \cdot L_e - G\sigma \sum_{f \in faces} \mathbf{F}_f \cdot \omega_f . \quad (19)$$

where G=$6.67 \times 10^{-11}$ m$^3$kg$^{-1}$s$^{-2}$ is the gravitational constant, $\sigma$ is the density of the body;

$\mathbf{r}_e$ is a body-fixed vector from the field point to some fixed point on the edge e of face f, $\mathbf{r}_f$ is a body-fixed vector from the field point to any point in the face plane; $\mathbf{E}_e$ and $\mathbf{F}_f$ are defined in terms of face- and edge-normal vectors, $\mathbf{E}_e$ is the geometric parameter of edges while $\mathbf{F}_f$ is the geometric parameter of faces; $L_e$ is the factor of integration that operates over the space between the field point and edges e of faces f, $\omega_f$ is the signed solid angle subtended by planar region relative to the field point.

## 4.1 Periodic Orbits in the Potential Field of the Asteroid 6489 Golevka and the Comet 1P/ Halley

### 4.1.1 Asteroid 6489 Golevka

The estimated bulk density of asteroid 6489 Golevka is 2.7 $\mathrm{g\cdot cm^{-3}}$ [35], its rotational period is 6.026 h, and its overall dimensions are $0.35 \times 0.25 \times 0.25$ km [35]. Figure 3 show two periodic orbits around the asteroid 6489 Golevka, while Figure 4 shows the topological distribution of characteristic multipliers for these two periodic orbits. Figure 4 shows that these two periodic orbits belong to the periodic cases—Cases P4 and P5. Clearly, the two orbits have the same shape but belong to different cases, which implies that periodic orbits that belong to different cases might have the same shape and that the essential characteristic of periodic orbits is the topological type rather than shape. In addition, the number of shapes is infinite, while the number of different topological types is finite.

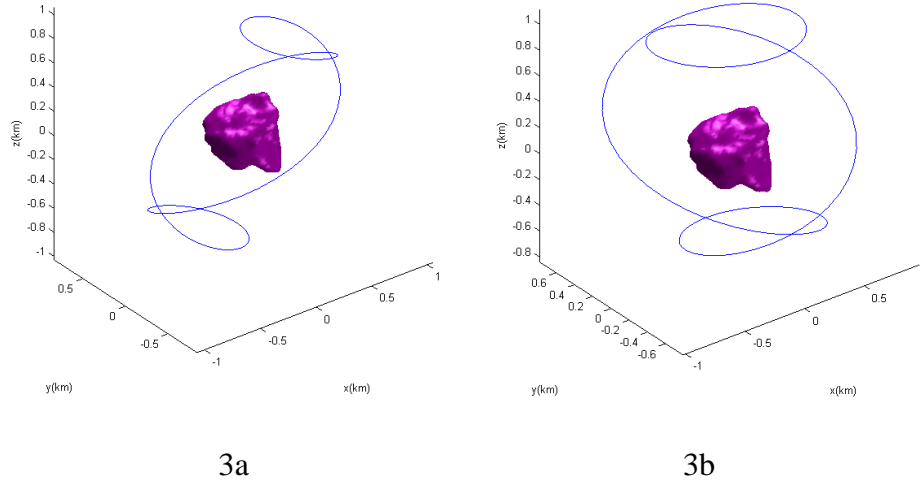

3a | 3b

Fig. 3. Two periodic orbits around the asteroid 6489 Golevka belong to Cases P4 and P5

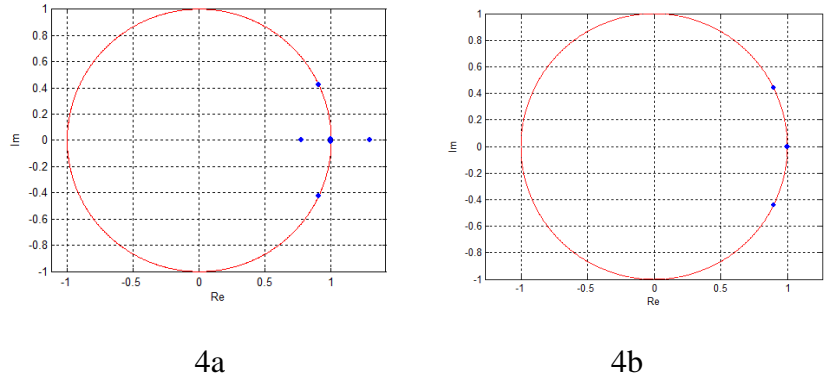

4a | 4b

Fig. 4. The topological distributions of characteristic multipliers for two periodic orbits around the asteroid 6489 Golevka

### 4.1.2 Comet 1P/ Halley

The nucleus of comet 1P/Halley is an irregular potato-shaped body [36]. The estimated bulk density of comet 1P/Halley's nucleus is 0.6 $g \cdot cm^{-3}$ [37], its rotational period is 52.8 h, and its overall dimensions are $16.831 \times 8.7674 \times 7.7692$ km [26][38].

Figure 5 show two periodic orbits around the comet 1P/Halley, while Figure 6 shows the topological distribution of characteristic multipliers for these two periodic orbits. Figure 6 shows that the periodic orbits in 5a belong to the stable periodic case, Case P2, while the periodic orbits in 5b belong to the unstable periodic case, Case P4, which means that different types of periodic orbits with different stabilities exist in the same highly irregular-shaped celestial body.

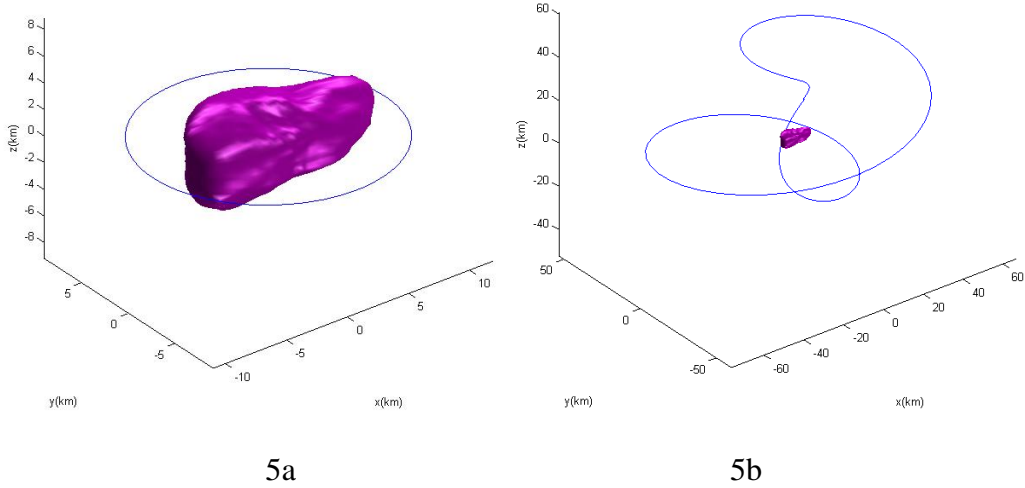

5a            5b

Fig. 5. Two periodic orbits around the comet 1P/Halley belong to Cases P2 and P4

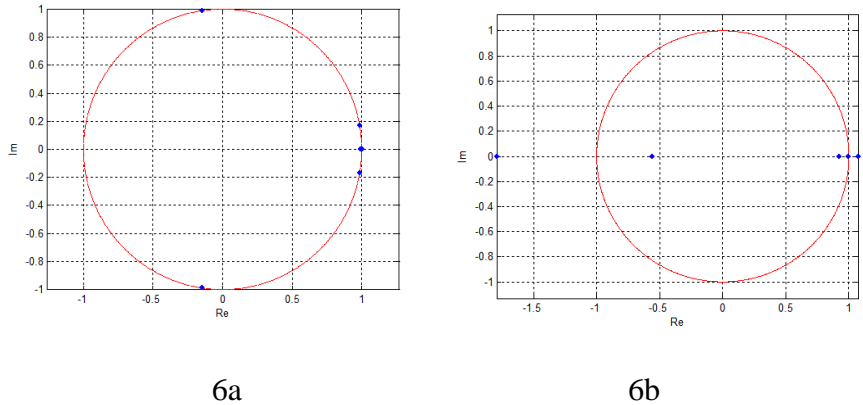

6a            6b

Fig. 6. The topological distributions of characteristic multipliers for these two periodic orbits around the comet 1P/Halley

## 4.2 Bifurcations of Periodic Orbits around Asteroid 216 Kleopatra

The estimated bulk density of asteroid 216 Kleopatra is 3.6 $\text{g}\cdot\text{cm}^{-3}$ [5], its rotational period is 5.385 h, and its overall dimensions are $217\times94\times81$ km [4]. The asteroid 216 Kleopatra is a large size ratio triple asteroid with two moonlets, which are Alexhelios (S/2008 (216) 1) and Cleoselene (S/2008 (216) 2) [5]. Alexhelios is the outer moonlet, with an estimated size of 8.9 ± 1.6 km, while Cleoselene is the inner moonlet, with an estimated size of 6.9 ± 1.6 km [5].

The study of dynamical behaviour around asteroid 216 Kleopatra can help explain the dynamical configuration of this large size ratio triple asteroid, including the orbital evolution, the motion stability, and the regular and chaotic motion of Alexhelios (S/2008 (216) 1) and Cleoselene (S/2008 (216) 2). Yu and Baoyin [20] found four equilibrium points in the potential of asteroid 216 Kleopatra, and all of these four equilibrium points are unstable [18][20]. Figure 7 shows the contour line of the zero-velocity manifold in the xy-plane, which shows that there are at least four equilibrium points outside the asteroid 216 Kleopatra. The hierarchical grid searching method can be used to search periodic orbits around irregular bodies; for 216 Kleopatra, this searching method was used to generate 29 basic periodic orbit families [19]. Collisions [39], perturbations from the sun [40], and the Yarkovsky effect [41] can lead to disruption of the asteroid [42] through bifurcation of the rotational stability and orbital motion [40][43]. Figure 7 clearly shows that there are 4 critical points outside this asteroid [18][20]. Additionally, from figure 7, one can see that there is a critical point near the origin of the body-fixed frame, which is unstable and implies that asteroid 216 Kleopatra will split near the unstable critical point inside the body.

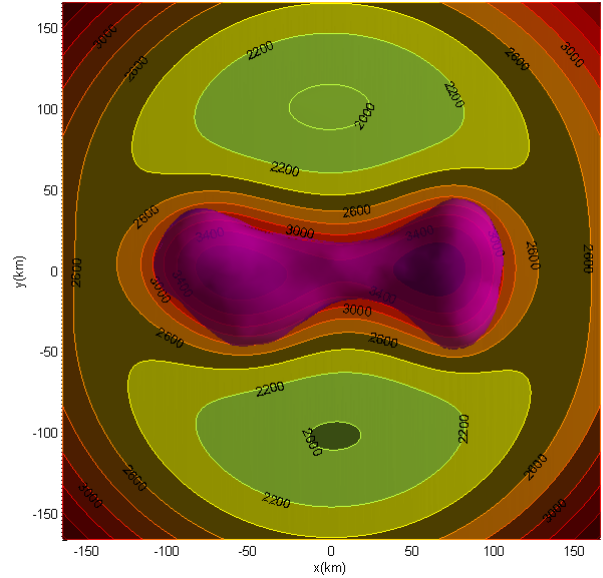

Fig. 7. The contour line of the zero-velocity manifold for asteroid 216 Kleopatra in the xy-plane. (The unit of the effective potential is m/s.)

## 4.2.1 Period-doubling bifurcations of periodic orbits around asteroid 216 Kleopatra

The period-doubling bifurcation of periodic orbits around asteroid 216 Kleopatra is discovered and shown in figure 8. Considering the Jacobi integral as the parameter, the periodic orbits change if the parameter changes; there are four periodic orbits around asteroid 216 Kleopatra presented in figure 8, with Jacobi integrals equal to 0.3135, 0.5635, 0.8135, and 1.0635. The unit of the Jacobi integral throughout this paper is $10^3 \text{ m}^2\text{s}^{-2}$. Figure 8b shows that the motion of characteristic multipliers belongs to the period-doubling bifurcation: Case P3 $\rightarrow$ Case PPD3 $\rightarrow$ Case P4.

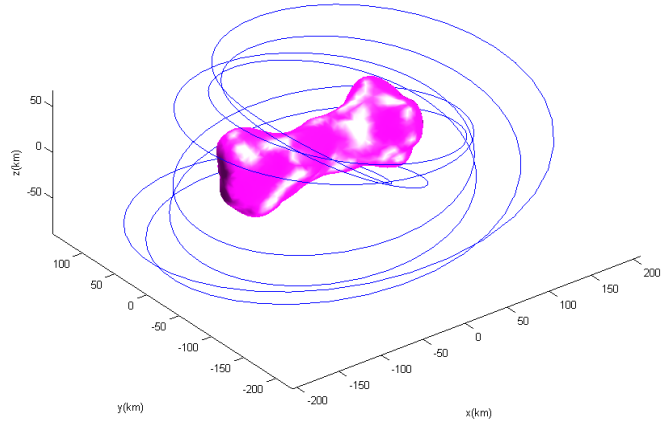

Fig. 8a. Four periodic orbits around asteroid 216 Kleopatra when the Jacobi integral parameter changes. The Jacobi integral is equal to 0.3135, 0.5635, 0.8135, and 1.0635, with a unit of $10^3 \text{ m}^2\text{s}^{-2}$.

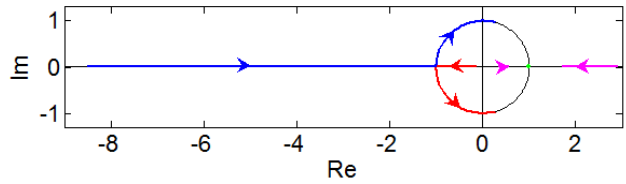

Fig. 8b. Motion of characteristic multipliers that leads to the period-doubling bifurcation
Case P3 $\rightarrow$ Case PPD3 $\rightarrow$ Case P4

### 4.2.2 Tangent bifurcations of periodic orbits around asteroid 216 Kleopatra

The tangent bifurcation of periodic orbits around asteroid 216 Kleopatra is shown in figure 9. The Jacobi integral is also considered as the parameter. Three periodic orbits around asteroid 216 Kleopatra are presented in figure 9, with the Jacobi integral equal to 1.0635, 1.3135, and 1.5635. From figure 9b, one can see that the motion of characteristic multipliers belongs to the tangent bifurcation: Case P4 $\rightarrow$ Case P5 $\rightarrow$ Case P2. The periodic orbit with the Jacobi integral 1.0635 belongs to Case P4, while the periodic orbits with the Jacobi integrals 1.3135 and 1.5635 belong to Case P2.

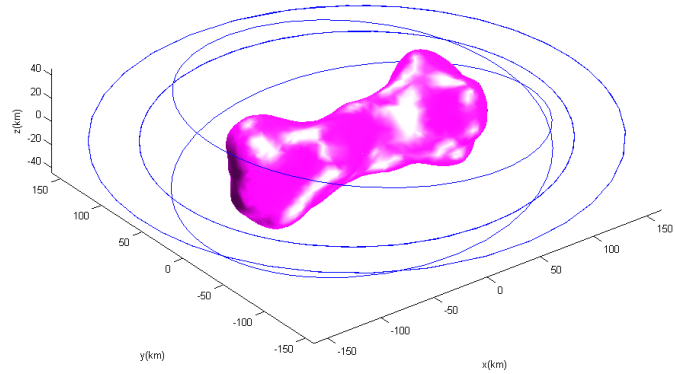

Fig. 9a. Three periodic orbits around asteroid 216 Kleopatra when the Jacobi integral parameter changes. The Jacobi integral equals 1.0635, 1.3135, and 1.5635, with a unit of $10^3 \text{ m}^2\text{s}^{-2}$.

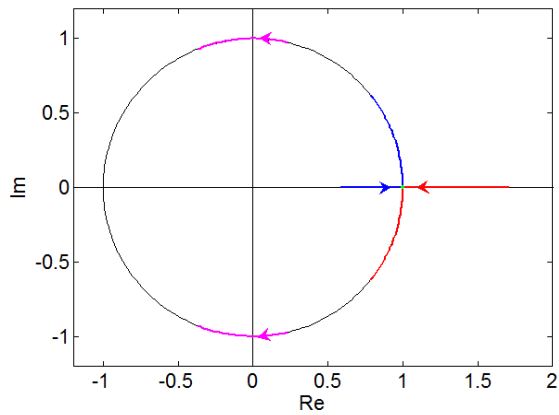

Fig. 9b. Motion of characteristic multipliers that leads to the tangent bifurcation
Case P4 $\rightarrow$ Case P5 $\rightarrow$ Case P2

## 4.2.3 Neimark-Sacker bifurcations of periodic orbits around asteroid 216 Kleopatra

The Neimark-Sacker bifurcation of periodic orbits around asteroid 216 Kleopatra is shown in figure 10. Three periodic orbits around asteroid 216 Kleopatra are presented in figure 10, with the Jacobi integral equal to 0.8392569, 0.9392569, and 1.0392569. The motion of the characteristic multipliers belongs to the Neimark-Sacker bifurcation: Case P2 $\rightarrow$ Case PK1 $\rightarrow$ Case P1. The periodic orbits with the Jacobi integrals 0.8392569 and 0.9392569 belong to Case P2, while the periodic orbit with the Jacobi integral 1.0392569 belongs to Case P1.

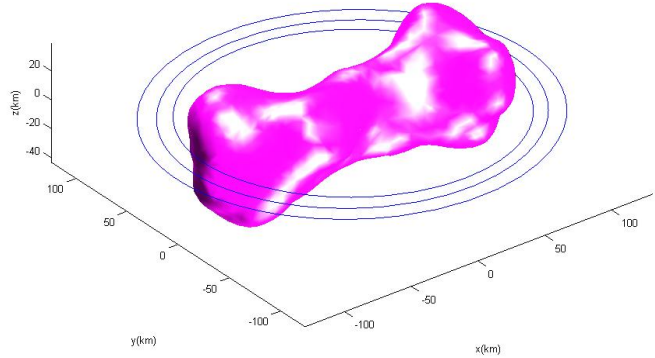

Fig. 10a. Three periodic orbits around asteroid 216 Kleopatra when the Jacobi integral parameter changes. The Jacobi integral equals 0.8392569, 0.9392569, and 1.0392569, with a unit of $10^3 \text{ m}^2\text{s}^{-2}$.

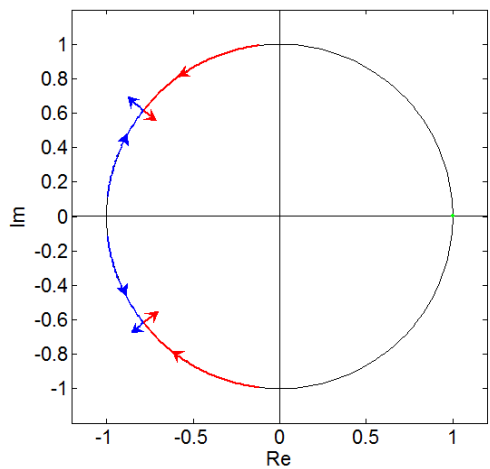

Fig. 10b. Motion of characteristic multipliers that leads to the Neimark-Sacker bifurcation
Case P2 $\to$ Case PK1 $\to$ Case P1

## 4.2.4 Real saddle bifurcations of periodic orbits around asteroid 216 Kleopatra

The real saddle bifurcation of periodic orbits around asteroid 216 Kleopatra is shown in figure 11. Four periodic orbits around asteroid 216 Kleopatra are presented in figure 11, with the Jacobi integral equal to -1.880585, -1.780585, -1.680585, and -1.580585. The motion of characteristic multipliers belongs to the real saddle bifurcation: Case P2 $\to$ Case PDRS1 $\to$ Case P1. The periodic orbits with the Jacobi integrals -1.880585 and -1.780585 belong to Case P2, while the periodic orbits with the Jacobi integrals -1.680585 and -1.580585 belong to Case P1. Figures 11a shows the attraction and exclusion effect of the critical points (Jiang et al. 2014[19]).

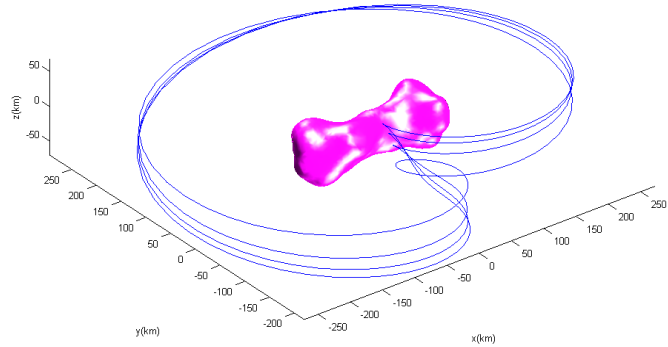

Fig. 11a. Four periodic orbits around asteroid 216 Kleopatra when the Jacobi integral parameter changes. The Jacobi integral equals -1.880585, -1.780585, -1.680585, and -1.580585, with a unit of $10^3 \text{ m}^2\text{s}^{-2}$.

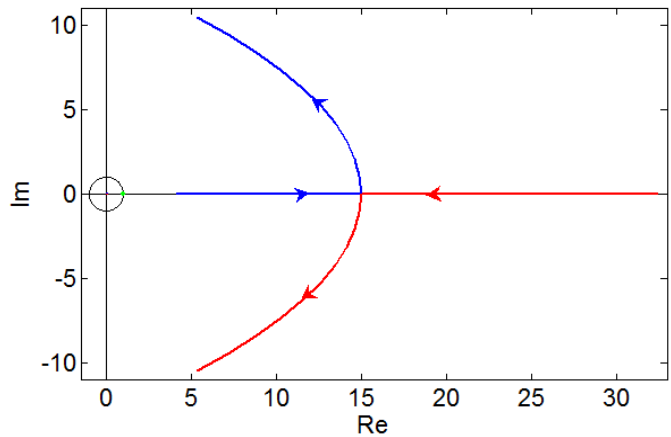

Fig. 11b. Motion of characteristic multipliers that leads to the real saddle bifurcation
Case P2 $\rightarrow$ Case PDRS1 $\rightarrow$ Case P1

## 4.3 Discussion of Applications

The theory developed in the previous sections has been applied to asteroids 216 Kleopatra and 6489 Golevka, as well as the comet nucleus of 1P/Halley. Two periodic orbits that have similar shapes and belong to different cases around the asteroid 6489 Golevka are shown; this result implies that periodic orbits that belong to the different cases might have similar shapes and that the essential characteristic of periodic orbits is the topological type rather than shape. For the comet 1P/Halley, two periodic orbits are found. These two periodic orbits belong to different periodic cases: one belongs to Case P2, the stable periodic case, while the other belongs to Case P4, the unstable periodic case.

Dynamical behaviours around the large size ratio triple asteroid 216 Kleopatra are presented in detail. There is an unstable critical point near the origin of the body-fixed frame for asteroid 216 Kleopatra, and asteroid 216 Kleopatra will split near this unstable critical point. All four types of bifurcations forecasted by the theory in this paper, specifically tangent bifurcations, period-doubling bifurcations, Neimark-Sacker bifurcations and real saddle bifurcations of periodic orbits, are discovered in the potential field of asteroid 216 Kleopatra. The parameter is the Jacobi integral. Each type of bifurcation is shown with the refinement of the parametric variation.

## 5. Conclusions

Nonlinearly dynamical behaviours in the potential field of an irregular body are discussed in this paper. Different distributions of characteristic multipliers are shown to fix the structure of submanifolds, the type of orbits, the dynamical behaviour, and the phase diagram of the motion. The topological classifications and stabilities of orbits in the potential field of an irregular body are presented. The classification includes 34 cases: 6 ordinary cases, 3 collisional cases, 3 degenerate real saddle cases, 7 periodic cases, 7 period-doubling cases, 1 periodic and collisional case, 1 and degenerate real saddle case, 1 period-doubling and collisional case, 1 period-doubling and degenerate real saddle case, and 4 periodic and period-doubling cases.

For periodic orbits, the period-doubling bifurcation, the tangent bifurcation, the Neimark-Sacker bifurcation, and the real saddle bifurcation of periodic orbits are discovered with parametric variation. Motions near bifurcation are sensitive to initial conditions. It is found that submanifolds appear to be Mobius strips and Klein bottles when the period-doubling bifurcation occurs.

As an application of the theory developed here, we study the relevant periodic orbits for the asteroids 216 Kleopatra and 6489 Golevka and the comet 1P/Halley. From the application to asteroid 6489 Golevka, the periodic orbits that belong to different cases can be shown to have potentially similar shapes, and the essential characteristic of periodic orbits is the topological type rather than shape. The application to comet 1P/Halley implies that different types of periodic orbits with different stabilities exist in the same highly irregular-shaped celestial body. All four types of bifurcations forecasted by the theory in this paper are discovered around asteroid 216 Kleopatra, which implies the complex dynamical behaviour in the potential field of this large size ratio triple asteroid.

## Acknowledgements

This research was supported by the National Basic Research Program of China (973 Program, 2012CB720000), the State Key Laboratory Foundation of Astronautic Dynamics (No. 2014ADL0202), and the National Natural Science Foundation of China (No. 11372150).

## Appendix 1

Table 1.a Classifications and properties for the ordinary cases

(C1: Cases; C2: Stability; C3: Phase diagram of motion; S: Stable; U: Unstable; K: Krein collision)

| C1 | Characteristic multipliers | C2 | C3 |
|---|---|---|---|
| O1 | $e^{\pm i\beta_j}\begin{pmatrix}\beta_j \in (0,\pi); j=1,2,3\\ \|\forall k \neq j, k=1,2,3, s.t. \beta_k \neq \beta_j\end{pmatrix}$ | S | Elliptic point |
| O2 | $\mathrm{sgn}(\alpha_j)e^{\pm\alpha_j}\left(\alpha_j \in \mathrm{R}, \|\alpha_j\| \in (0,1), j=1\right)$<br>$e^{\pm i\beta_j}\left(\beta_j \in (0,\pi); j=1,2 \| \beta_1 \neq \beta_2\right)$ | U | Mixed point: Elliptic point in $W^c(\mathbf{S})$;<br>Exponential saddle in $\bar{W}^s(\mathbf{S}) \oplus \bar{W}^u(\mathbf{S})$ |
| O3 | $\mathrm{sgn}(\alpha_j)e^{\pm\alpha_j}\begin{pmatrix}\alpha_j \in \mathrm{R}, \|\alpha_j\| \in (0,1);\\ j=1,2\|\alpha_1 \neq \alpha_2\end{pmatrix}$<br>$e^{\pm i\beta_j}\left(\beta_j \in (0,\pi), j=1\right)$ | U | Mixed point: Elliptic point in $W^c(\mathbf{S})$;<br>Exponential saddle in $\bar{W}^s(\mathbf{S}) \oplus \bar{W}^u(\mathbf{S})$ |
| O4 | $e^{\pm i\beta_j}\left(\beta_j \in (0,\pi), j=1\right)$<br>$e^{\pm\sigma \pm i\tau}\left(\sigma,\tau \in \mathrm{R}; \sigma>0, \tau \in (0,\pi)\right)$ | U | Mixed point:<br>Elliptic point in $W^c(\mathbf{S})$;<br>Spiral saddle in $\tilde{W}^s(\mathbf{S}) \oplus \tilde{W}^u(\mathbf{S})$ |
| O5 | $\mathrm{sgn}(\alpha_j)e^{\pm\alpha_j}\left(\alpha_j \in \mathrm{R}, \|\alpha_j\| \in (0,1), j=1\right)$<br>$e^{\pm\sigma \pm i\tau}\left(\sigma,\tau \in \mathrm{R}; \sigma>0, \tau \in (0,\pi)\right)$ | U | Mixed point: Exponential saddle in $\bar{W}^s(\mathbf{S}) \oplus \bar{W}^u(\mathbf{S})$;<br>Spiral saddle in $\tilde{W}^s(\mathbf{S}) \oplus \tilde{W}^u(\mathbf{S})$ |
| O6 | $\mathrm{sgn}(\alpha_j)e^{\pm\alpha_j}\begin{pmatrix}\alpha_j \in \mathrm{R}, \|\alpha_j\| \in (0,1), j=1,2,3\\ \|\forall k \neq j, k=1,2,3, s.t. \alpha_k \neq \alpha_j\end{pmatrix}$ | U | Exponential saddle |

Table 1.b Classifications and properties for the periodic cases

(C1: Cases; C2: Stability; C3: Phase diagram of motion; S: Stable; U: Unstable; K: Krein collision)

| C1 | Characteristic multipliers | C2 | C3 |
|---|---|---|---|
| P1 | $\gamma_j \left(\gamma_j = 1; j=1,2\right)$<br>$e^{\pm\sigma \pm i\tau}\left(\sigma,\tau \in \mathrm{R}; \sigma>0, \tau \in (0,\pi)\right)$ | U | Mixed point:<br>Collisional in $W^e(\mathbf{S})$;<br>Spiral saddle in $\tilde{W}^s(\mathbf{S}) \oplus \tilde{W}^u(\mathbf{S})$ |
| P2 | $\gamma_j \left(\gamma_j = 1; j=1,2\right)$<br>$e^{\pm i\beta_j}\left(\beta_j \in (0,\pi); j=1,2\|\beta_1 \neq \beta_2\right)$ | K | Mixed point:<br>Collisional in $W^e(\mathbf{S})$;<br>Elliptic point in $W^c(\mathbf{S})$ |
| P3 | $\gamma_j \left(\gamma_j = 1; j=1,2\right)$<br>$\mathrm{sgn}(\alpha_j)e^{\pm\alpha_j}\begin{pmatrix}\alpha_j \in \mathrm{R}, \|\alpha_j\| \in (0,1);\\ j=1,2\|\alpha_1 \neq \alpha_2\end{pmatrix}$ | K | Mixed point: Collisional in $W^e(\mathbf{S})$;<br>Exponential saddle in $\bar{W}^s(\mathbf{S}) \oplus \bar{W}^u(\mathbf{S})$ |
| P4 | $\gamma_j \left(\gamma_j = 1; j = 1, 2\right)$ $e^{\pm i\beta_j} \left(\beta_j \in (0, \pi), j = 1\right)$ $\text{sgn}(\alpha_j) e^{\pm \alpha_j} \left(\alpha_j \in \mathbb{R}, |\alpha_j| \in (0,1), j = 1\right)$ | K | Mixed point: Collisional in $W^e(\mathbf{S})$; Elliptic point in $W^c(\mathbf{S})$; Exponential saddle in $\bar{W}^s(\mathbf{S}) \oplus \bar{W}^u(\mathbf{S})$ |
| P5 | $\gamma_j \left(\gamma_j = 1; j = 1,2,3,4\right)$ $e^{\pm i\beta_j} \left(\beta_j \in (0, \pi), j = 1\right)$ | K | Mixed point: Collisional in $W^e(\mathbf{S})$; Elliptic point in $W^c(\mathbf{S})$ |
| P6 | $\gamma_j \left(\gamma_j = 1; j = 1,2,3,4\right)$ $\text{sgn}(\alpha_j) e^{\pm \alpha_j} \left(\alpha_j \in \mathbb{R}, |\alpha_j| \in (0,1), j = 1\right)$ | K | Mixed point: Collisional in $W^e(\mathbf{S})$; Exponential saddle in $\bar{W}^s(\mathbf{S}) \oplus \bar{W}^u(\mathbf{S})$ |
| P7 | $\gamma_j \left(\gamma_j = 1; j = 1,2,...,6\right)$ | K | Collisional point |

Table 1.c Classifications and properties for the period-doubling cases

(C1: Cases; C2: Stability; C3: Phase diagram of motion; S: Stable; U: Unstable; K: Krein collision)

| C1 | Characteristic multipliers | C2 | C3 |
|---|---|---|---|
| PD1 | $\gamma_j \left(\gamma_j = -1; j = 1,2\right)$ $e^{\pm \sigma \pm i\tau} \left(\sigma, \tau \in \mathbb{R}; \sigma > 0, \tau \in (0, \pi)\right)$ | U | Mixed point: Collisional in $W^e(\mathbf{S})$; Spiral saddle in $\tilde{W}^s(\mathbf{S}) \oplus \tilde{W}^u(\mathbf{S})$ |
| PD2 | $\gamma_j \left(\gamma_j = -1; j = 1,2\right)$ $e^{\pm i\beta_j} \left(\beta_j \in (0, \pi); j = 1,2 | \beta_1 \neq \beta_2\right)$ | K | Mixed point: Collisional in $W^e(\mathbf{S})$; Elliptic point in $W^c(\mathbf{S})$ |
| PD3 | $\gamma_j \left(\gamma_j = -1; j = 1,2\right)$ $\text{sgn}(\alpha_j) e^{\pm \alpha_j} \begin{pmatrix} \alpha_j \in \mathbb{R}, |\alpha_j| \in (0,1); \\ j = 1,2 | \alpha_1 \neq \alpha_2 \end{pmatrix}$ | K | Mixed point: Collisional in $W^e(\mathbf{S})$; Degenerate exponential saddle in $\bar{W}^s(\mathbf{S}) \oplus \bar{W}^u(\mathbf{S})$ |
| PD4 | $\gamma_j \left(\gamma_j = -1; j = 1,2\right)$ $e^{\pm i\beta_j} \left(\beta_j \in (0, \pi), j = 1\right)$ $\text{sgn}(\alpha_j) e^{\pm \alpha_j} \left(\alpha_j \in \mathbb{R}, |\alpha_j| \in (0,1), j = 1\right)$ | K | Mixed point: Collisional in $W^e(\mathbf{S})$; Elliptic point in $W^c(\mathbf{S})$; Exponential saddle in $\bar{W}^s(\mathbf{S}) \oplus \bar{W}^u(\mathbf{S})$ |
| PD5 | $\gamma_j \left(\gamma_j = -1; j = 1,2,3,4\right)$ | K | Mixed point: Collisional in $W^e(\mathbf{S})$; |

|  | $e^{\pm i\beta_j}\left(\beta_j \in (0,\pi), j=1\right)$ |  | Elliptic point in $W^c(\mathbf{S})$ |
| --- | --- | --- | --- |
| PD6 | $\gamma_j \left(\gamma_j = -1; j=1,2,3,4\right)$ | K | Mixed point: Collisional in $W^e(\mathbf{S})$; Exponential saddle in |
|  | $\text{sgn}(\alpha_j)e^{\pm\alpha_j}\left(\alpha_j \in \mathrm{R}, |\alpha_j| \in (0,1), j=1\right)$ |  | $\bar{W}^s(\mathbf{S}) \oplus \bar{W}^u(\mathbf{S})$ |
| PD7 | $\gamma_j \left(\gamma_j = -1; j=1,2,\ldots,6\right)$ | K | Collisional point |

Table 1.d Classifications and properties for the collisional cases

(C1: Cases; C2: Stability; C3: Phase diagram of motion; S: Stable; U: Unstable; K: Krein collision)

| C1 | Characteristic multipliers | C2 | C3 |
| --- | --- | --- | --- |
| K1 | $e^{\pm i\beta_j}\left(\beta_j \in (0,\pi), \beta_1 = \beta_2 = \beta_3; j=1,2,3\right)$ | K | Collisional point |
| K2 | $e^{\pm i\beta_j}\left(\beta_j \in (0,\pi), \beta_1 = \beta_2 \neq \beta_3; j=1,2,3\right)$ | K | Collisional point |
| K3 | $\text{sgn}(\alpha_j)e^{\pm\alpha_j}\left(\alpha_j \in \mathrm{R}, |\alpha_j| \in (0,1), j=1\right)$ $e^{\pm i\beta_j}\left(\beta_j \in (0,\pi), \beta_1 = \beta_2; j=1,2\right)$ | K | Collisional point |

Table 1.e Classifications and properties for the degenerate real saddle cases

(C1: Cases; C2: Stability; C3: Phase diagram of motion; S: Stable; U: Unstable; K: Krein collision)

| C1 | Characteristic multipliers | C2 | C3 |
| --- | --- | --- | --- |
| DRS1 | $\text{sgn}(\alpha_j)e^{\pm\alpha_j}\begin{pmatrix}\alpha_j \in \mathrm{R}, |\alpha_j| \in (0,1), j=1,2,3\\ \alpha_1 = \alpha_2 \neq \alpha_3\end{pmatrix}$ | U | Degenerate exponential saddle |
| DRS2 | $\text{sgn}(\alpha_j)e^{\pm\alpha_j}\begin{pmatrix}\alpha_j \in \mathrm{R}, |\alpha_j| \in (0,1), j=1,2,3\\ \alpha_1 = \alpha_2 = \alpha_3\end{pmatrix}$ | U | Degenerate exponential saddle |
| DRS3 | $e^{\pm i\beta_j}\left(\beta_j \in (0,\pi); j=1\right)$ $\text{sgn}(\alpha_j)e^{\pm\alpha_j}\begin{pmatrix}\alpha_j \in \mathrm{R}, |\alpha_j| \in (0,1), j=1,2\\ \alpha_1 = \alpha_2\end{pmatrix}$ | U | Mixed point: Elliptic point in $W^c(\mathbf{S})$; Degenerate exponential saddle in $\bar{W}^s(\mathbf{S}) \oplus \bar{W}^u(\mathbf{S})$ |

Table 1.f Classifications and properties for the mixed cases

(C1: Cases; C2: Stability; C3: Phase diagram of motion; S: Stable; U: Unstable; K: Krein collision)

| C1 | Characteristic multipliers | C2 | C3 |
| --- | --- | --- | --- |
| PK1 | $\gamma_j \ (\gamma_j = 1; j = 1,2)$ $e^{\pm i\beta_j} \ (\beta_j \in (0,\pi), \beta_1 = \beta_2; j = 1,2)$ | S | Mixed point: Fixed point in $W^e(\mathbf{S})$; Elliptic point in $W^c(\mathbf{S})$ |
| PDRS1 | $\gamma_j \ (\gamma_j = 1; j = 1,2)$ $\text{sgn}(\alpha_j) e^{\pm \alpha_j} \begin{pmatrix} \alpha_j \in \mathrm{R}, |\alpha_j| \in (0,1), j = 1,2 \\ \alpha_1 = \alpha_2 \end{pmatrix}$ | U | Mixed point: Fixed point in $W^e(\mathbf{S})$; Degenerate exponential saddle in $\bar{W}^s(\mathbf{S}) \oplus \bar{W}^u(\mathbf{S})$ |
| PDK1 | $\gamma_j \ (\gamma_j = -1; j = 1,2)$ $e^{\pm i\beta_j} \ (\beta_j \in (0,\pi), \beta_1 = \beta_2; j = 1,2)$ | S | Mixed point: Flipping point in $W^d(\mathbf{S})$; Elliptic point in $W^c(\mathbf{S})$ |
| PDDRS1 | $\gamma_j \ (\gamma_j = -1; j = 1,2)$ $\text{sgn}(\alpha_j) e^{\pm \alpha_j} \begin{pmatrix} \alpha_j \in \mathrm{R}, |\alpha_j| \in (0,1), j = 1,2 \\ \alpha_1 = \alpha_2 \end{pmatrix}$ | U | Mixed point: Flipping point in $W^d(\mathbf{S})$; Degenerate exponential saddle in $\bar{W}^s(\mathbf{S}) \oplus \bar{W}^u(\mathbf{S})$ |
| PPD1 | $\gamma_j \ (\gamma_j = 1; j = 1,2,3,4)$ $\gamma_j \ (\gamma_j = -1; j = 5,6)$ | S | Mixed point: Fixed point in $W^e(\mathbf{S})$; Flipping point in $W^d(\mathbf{S})$ |
| PPD2 | $\gamma_j \ (\gamma_j = -1; j = 1,2,3,4)$ $\gamma_j \ (\gamma_j = 1; j = 5,6)$ | S | Mixed point: Fixed point in $W^e(\mathbf{S})$; Flipping point in $W^d(\mathbf{S})$ |
| PPD3 | $\gamma_j \ (\gamma_j = -1; j = 1,2)$ $\gamma_j \ (\gamma_j = 1; j = 3,4)$ $e^{\pm i\beta_j} \ (\beta_j \in (0,\pi), j = 1)$ | S | Mixed point: Fixed point in $W^e(\mathbf{S})$; Flipping point in $W^d(\mathbf{S})$; Elliptic point in $W^c(\mathbf{S})$ |
| PPD4 | $\gamma_j \ (\gamma_j = -1; j = 1,2)$ $\gamma_j \ (\gamma_j = 1; j = 3,4)$ $\text{sgn}(\alpha_j) e^{\pm \alpha_j} \ (\alpha_j \in \mathrm{R}, |\alpha_j| \in (0,1), j = 1)$ | U | Mixed point: Fixed point in $W^e(\mathbf{S})$; Flipping point in $W^d(\mathbf{S})$; Degenerate exponential saddle in $\bar{W}^s(\mathbf{S}) \oplus \bar{W}^u(\mathbf{S})$ |